\documentclass[conference]{IEEEtran}
\IEEEoverridecommandlockouts
\usepackage{amsmath,amssymb,amsfonts}
\usepackage{algorithmic}
\usepackage{graphicx}
\usepackage{textcomp}
\usepackage{xcolor}
\usepackage{bigints}
\usepackage{subcaption}
\usepackage{float}
\usepackage[hidelinks]{hyperref} 
\usepackage[style=ieee]{biblatex}
\usepackage{svg}
\usepackage{balance}
\usepackage{multirow}
\usepackage{soul}
\usepackage{xcolor}
\usepackage{diagbox}
\usepackage{makecell}

\makeatletter 
\newcommand{\linebreakand}{%
  \end{@IEEEauthorhalign}
  \hfill\mbox{}\par
  \mbox{}\hfill\begin{@IEEEauthorhalign}
}

\def\BibTeX{{\rm B\kern-.05em{\sc i\kern-.025em b}\kern-.08em
    T\kern-.1667em\lower.7ex\hbox{E}\kern-.125emX}}
\begin{document}

\title{
Introducing Combined Effects of Filtering and ASE Noise in Optical Links Supposing Different Equalization Algorithms
\\
{}
}

\author{\IEEEauthorblockN{1\textsuperscript{st} Enrico Miotto}
\IEEEauthorblockA{
\textit{DET, Politecnico di Torino}\\
Turin, Italy\\
enrico.miotto@polito.it}
\and
\IEEEauthorblockN{2\textsuperscript{nd} Andrea Rosso}
\IEEEauthorblockA{
\textit{DET, Politecnico di Torino}\\
Turin, Italy\\
andrea.rosso@polito.it}
\and
\IEEEauthorblockN{3\textsuperscript{rd} Emanuele Virgillito}
\IEEEauthorblockA{
\textit{DET, Politecnico di Torino}\\
Turin, Italy}
\and
\IEEEauthorblockN{4\textsuperscript{th} Stefano Straullu}
\IEEEauthorblockA{\textit{LINKS Foundation} \\
Turin, Italy}
\and

\linebreakand

\IEEEauthorblockN{5\textsuperscript{th} Andrea Castoldi}
\IEEEauthorblockA{\textit{SM-Optics} \\
Cologno Monzese, Italy\\
}
\and
\IEEEauthorblockN{6\textsuperscript{th} Andrea Bovio}
\IEEEauthorblockA{\textit{SM-Optics} \\
Cologno Monzese, Italy\\
}
\and
\IEEEauthorblockN{7\textsuperscript{th} Francisco Martinez Rodriguez}
\IEEEauthorblockA{\textit{SM-Optics} \\
Cologno Monzese, Italy\\
}
\and
\IEEEauthorblockN{8\textsuperscript{th} Vittorio Curri}
\IEEEauthorblockA{ \textit{DET, Politecnico di Torino}\\
Turin, Italy}
}

\maketitle

\begin{abstract} 
This paper develops and validates a discrete-time modeling framework for the joint impact of cascaded optical filtering, distributed ASE noise, and transceiver noise in coherent optical links. The work focuses on the post-equalization signal-to-noise ratio, which is the central quantity used to quantify filtering penalty under receiver DSP. Starting from an optical-link abstraction with arbitrary filter transfer functions and colored noise spectra, we derive analytical expressions for the matched-filter bound, Zero-Forcing Equalization, Minimum Mean Square Error Equalization, Fractionally Spaced Equalization, and finite-length equalizers with and without explicit colored-noise treatment. The model is coupled to a measurement-based transceiver SNR characterization, so that optical-link penalties and receiver impairments can be evaluated within the same formulation. Time-domain simulations with LMS equalization validate the analytical predictions over severe filtering conditions, different tap lengths, and different ASE-noise positions along the link. Experimental results with commercial transceivers and ROADMs further confirm the accuracy of the MMSE and FSE models, while highlighting the role of realistic filter modeling and equalizer implementation limits. The resulting framework provides a tractable basis for quality-of-transmission estimation and optical-network digital-twin implementations.
\end{abstract}

\begin{IEEEkeywords}
filtering penalty, equalization, MMSE, mathematical modeling
\end{IEEEkeywords}

\section{Introduction}

Metro and access optical networks are increasingly required to support high-capacity coherent transmission while keeping quality-of-transmission estimates accurate and computationally efficient. In these scenarios, linear impairments often dominate over nonlinear propagation effects, because links are shorter and the launched optical powers are generally lower than in long-haul systems\cite{roadmap}. Among the relevant linear impairments, cascaded optical filtering is especially important: each ROADM or optical filtering element reshapes both the useful signal and the noise already present in the line, producing a filtering penalty that depends on the full sequence of filters, noise sources, and receiver equalization strategy \cite{Delezoide2018}.

This issue is also central for optical network digital twins (DTs). Accurate fiber and amplifier models, including transparent-lightpath approximations based on the Gaussian Model, have enabled increasingly reliable system-level estimators \cite{gnpy}. However, filtering effects are still difficult to include in a disaggregated and mathematically consistent way, especially when the receiver DSP and the relative position of ASE-noise injection along the lightpath are relevant. The objective of this work is therefore to provide a modeling framework in which cascaded filtering, distributed ASE noise, transceiver impairment, and equalization are treated within a single SNR-based formulation.

The core of the paper is the derivation and validation of post-equalization SNR models for coherent systems affected by optical filtering. We first introduce a transceiver model that maps experimental back-to-back BER measurements into an equivalent transceiver SNR, following the approach of \cite{Toru2024, TesiAndrea} and its use in \cite{ONDM2025}. This transceiver contribution is then combined with the ASE contribution in the optical-link abstraction used for the analytical derivations. Since nonlinear interference is neglected in the target metro-access use case \cite{roadmap}, the analysis isolates the effects of filtering and additive noise.

Starting from this abstraction, we derive models for the Matched Filter Bound (MFB), Zero-Forcing Equalizer (ZFE), Minimum Mean Square Error Equalizer (MMSE), Fractionally Spaced Equalizer (FSE), and Finite Length Equalizer (FLE). The derivation is performed in discrete time, consistently with the operation of commercial coherent receivers after sampling. A key point of the model is that ASE contributions injected at different positions experience different downstream filters, while the transceiver noise is added at the receiver. As a consequence, the total noise at the equalizer input is generally colored, and the equalizer model must account for both signal distortion and noise enhancement.

The analytical models are validated in two complementary ways. First, time-domain simulation is used to compare the proposed formulas with LMS-based receiver equalization under controlled and severe filtering scenarios. This validation exploits the simulation framework already used in \cite{ECOC25} and allows the number of taps, the equivalent bandwidth of the filter cascade, and the position of the ASE-noise sources to be varied independently. Second, experimental measurements with commercial transceivers and accurately modeled optical filters, following the setup of \cite{ONDM2025}, are used to validate the MMSE and FSE predictions in realistic conditions.

The resulting framework clarifies the accuracy-complexity tradeoff among the considered equalizer models. Infinite-length models provide fast and useful estimates when the filtering condition is not extreme, while the finite-length formulation captures the limited-tap penalty observed in practical DSP. This makes the proposed approach suitable for quality-of-transmission estimation and for integration into optical-network DT transmission models, where filtering penalties must be evaluated accurately without relying on full time-domain simulation for every lightpath.

\section{Related Literature}
Equalization over bandwidth-limited channels is a classical topic in digital communications, and the main theoretical tools used in this paper originate from established treatments of linear channels, discrete-time representations, and MMSE-based receiver design \cite{Proakis2008}, \cite{cioffi},  \cite{cioffiD}. The optical-network case, however, requires a dedicated formulation. Unlike many wireless channels, the optical filtering transfer function can often be characterized or modeled with high accuracy, as discussed later in Section \ref{sec:filter_model}. At the same time, optical links include alternating amplification and filtering stages, so ASE noise is injected at multiple positions along the lightpath. Each ASE contribution then experiences a different downstream filtering response, and its spectrum at the receiver differs from both the transmitted signal spectrum and the other noise spectra. This produces colored noise, which is one of the main elements explicitly included in the derivations of Section \ref{sec:models}.

Previous experimental work has quantified filtering penalties and mitigation strategies in meshed optical networks with several ROADMs, including the use of optical wave shapers \cite{filt_mitigation}. Metro-access scenarios with a single ROADM and a single noise source have also been investigated experimentally \cite{rizzelli_metroaccess}. In addition, the influence of optical-filter frequency shifts and bandwidth variations on the Q-factor has been studied through experiments and simulations \cite{freq_band_exp}, showing that asymmetric and bandwidth-dependent filtering effects must be represented carefully when estimating system performance.

Several modeling approaches have addressed these effects. In \cite{nokia_zami_linearequal}, the degradation caused by in-line optical filtering in WDM wavelength-routing nodes is modeled using digital-communications theory for band-limited channels with linear equalization, and the approach is experimentally validated with a commercial transponder under nonlinear fiber transmission. The equalizer in that work is treated in continuous time through Fourier-transform analysis. A continuous-time equalizer is also considered in \cite{rizzelli}, where the model is experimentally validated on a setup with one ROADM and one ASE-noise source while accounting for filter-bandwidth variations and central-frequency shifts. More recently, \cite{Virgillito2024infinitelength} proposed an MMSE equalization model based on the DTFT, emphasizing the relevance of ASE/filter ordering and receiver-DSP implementation capabilities; in that case, bandwidth variation is also considered and validation is performed through numerical time-domain simulations.

The contribution of the present work is to extend this line of research toward the discrete-time receiver model that is most relevant for commercial coherent transceivers. The considered equalizers, including ZFE, MMSE, FSE, and FLE, operate after sampling, and the derivation admits arbitrary optical-filter shapes, arbitrary placement of ASE-noise sources, and colored transceiver noise. The transceiver impairment model is embedded in the same SNR formulation used for the optical link, so filtering penalties can be evaluated together with receiver limitations. Compared with models focused only on infinite-length equalization or on a single filtering/noise configuration, the proposed framework also includes finite-length equalizers and a simulation-based validation of the tap-limited regime. The complete set of assumptions specific to amplified optical links is stated explicitly, while the most technical passages are summarized and referenced inside the derivation.

\section{TRANSCEIVER MODEL AND SNR DEFINITION}
\label{sec:transceiver}

\begin{figure*}[htb!]
    \centering
    \includegraphics[width=0.7\linewidth]{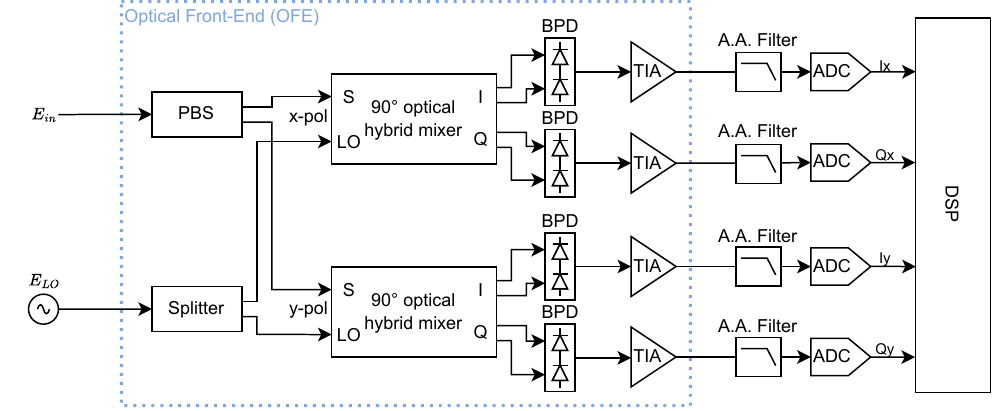}
    \caption{Realistic receiver block diagram. ADC: analog to digital converter; BPD: balanced photodetector; I: in-phase component; LO: local oscillator; PBS: polarization beam splitter;  Q: quadrature component;  TIA: transimpedance amplifier.}
    \label{fig:realistic_receiver}
\end{figure*}

\begin{figure}[ht]
    \centering
    \includegraphics[width=0.85\linewidth]{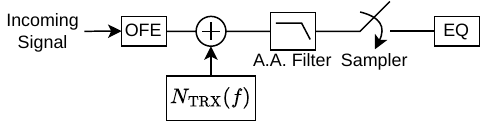}
    \caption{Receiver block diagram abstraction. AA: anti-aliasing filter; EQ: equalization block; OFE: optical front end block.}
    \label{fig:trx_block_diagram}
\end{figure}

This section introduces a measurement-based methodology for modeling TRX performance that accounts for transceiver impairments when evaluating the filtering penalty. The focus is on accurately modeling the SNR and its impact on the BER, particularly in scenarios dominated by linear impairments, such as metro networks. In this work, the experimental validation was performed using two commercial transceivers from two different vendors, referred to as Vendor A and Vendor B. For each transceiver, the three modulation formats considered and the corresponding symbol rates are reported in \autoref{table:symbol_rates}.

\begin{table}[tb!]
\centering
\begin{tabular}{|l|l|l|}
\hline
Modulation format & Vendor A & Vendor B \\ \hline
DP-QPSK 200G      & 63.1 GBd & 69.4 GBd \\ \hline
DP-16QAM 200G     & 31.6 GBd & 34.7 GBd \\ \hline
DP-16QAM 400G     & 63.1 GBd & 69.4 GBd \\ \hline
\end{tabular}
\caption{Symbol rate comparison for transceivers of different vendors for the considered modulation formats.}
\label{table:symbol_rates}
\end{table}

\autoref{fig:realistic_receiver} shows a coherent optical transceiver front-end in which the received signal, after random polarization rotations, is split into two orthogonal polarization branches by a polarization beam splitter. Each polarization is mixed with a local oscillator in a 90° optical hybrid, producing in-phase and quadrature components that are detected by balanced photodetectors. The resulting electrical currents are amplified by transimpedance amplifiers, passed through anti-aliasing (AA) filters, and digitized by parallel analog to digital converters (ADCs) for both x- and y-polarization states. The digital samples are then processed by the DSP stage to recover the transmitted information. 
To facilitate the analysis of filtering penalties, we adopt a simplified transceiver representation that preserves the impact of receiver impairments and the equalization strategy, while remaining tractable for system-level discussion.

The block diagram of the transceiver model is provided in \autoref{fig:trx_block_diagram}. The transceiver is represented by grouping the main physical and digital stages into equivalent functional blocks. The incoming signal is first processed by an equivalent ideal optical front-end (OFE), performing the optical to electrical conversion. The impairments of the realistic OFE of \autoref{fig:realistic_receiver} are summarized through an aggregated receiver noise contribution.
In addition, it is assumed that transmitter impairments are generally much smaller than receiver impairments, according to \cite{ONDM2025, miotto2025arxive, rizzelli_finitelength}. Therefore, all the transceiver noise is assumed to be concentrated on the receiver side.

The resulting electrical signal then passes through an AA filter that could introduce an overall analog bandwidth limitation before sampling. Finally, the sampled sequence is handled by an equivalent equalization stage that models the DSP action of the transceiver. This abstraction captures the dominant effects of noise, electrical filtering, and digital compensation while avoiding the complexity of the full schematic.

\begin{figure}[htb!]
    \centering
    \includegraphics[width=\linewidth]{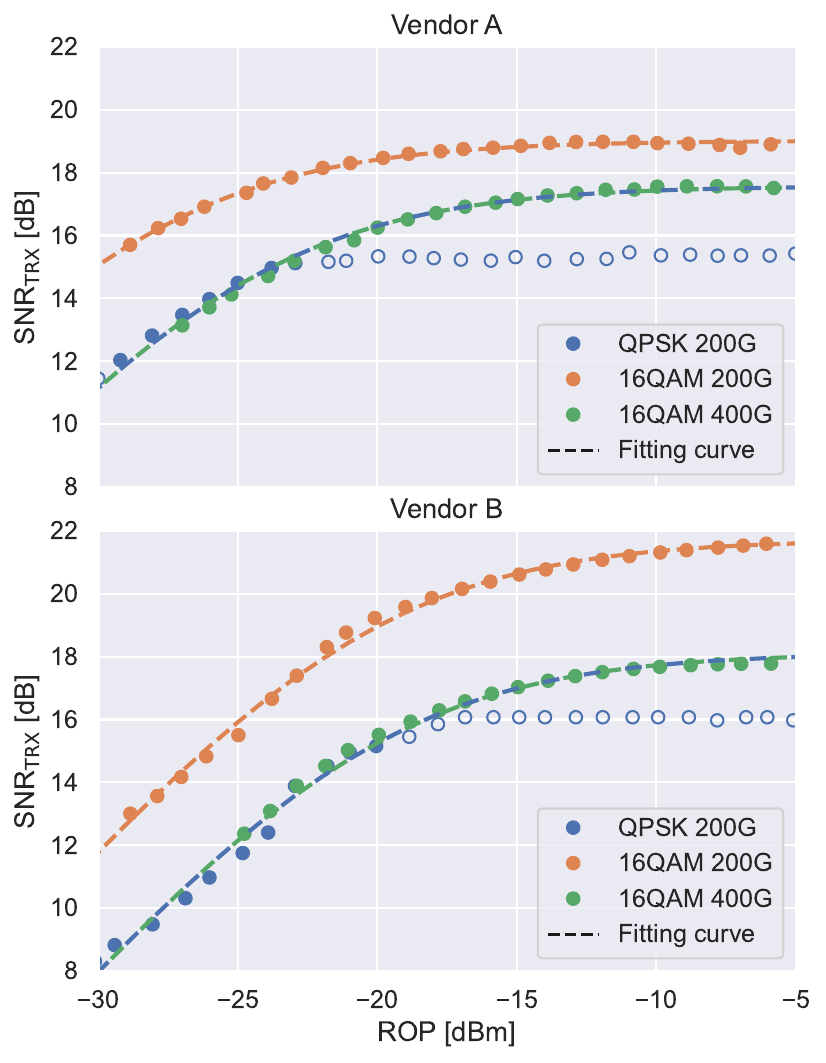}
    \caption{Back-to-back characterization of the transceivers.}
    \label{fig:transceiver}
\end{figure}

As a first approximation, the equivalent transceiver noise can be modeled as white, and the AA filter can be modeled as sufficiently broad so that no additional filtering effect is applied at the receiver stage.
For some commercial transceivers, it may be necessary to account for electrical filtering and colored noise in the receiver model, as will be discussed in later sections. In this section, we assess the transceiver performance for both vendors A and B, by exploiting a fitting formula for their equivalent SNR. The result of this procedure is provided in \autoref{fig:transceiver}.

The bit error rate is expressed as:
\begin{equation}
\text{BER} = k_1 \cdot \operatorname{erfc}\left( \sqrt{k_2 \cdot \text{SNR}}\right)
\label{eq:BER_vs_SNR}
\end{equation}
where $k_1$ and $k_2$ depend on the modulation format\cite{QAM,Carena:2012}. The total SNR of the system is obtained by considering the Gaussian Channel model, which has been shown to be applied in metro-access networks scenarios in \cite{RossoICTON:2025}, as the inverse sum of individual contributions:
\begin{equation}
\text{SNR}^{-1} = \text{SNR}_{\text{ASE}}^{-1} + \text{SNR}_{\text{TRX}}^{-1}
\label{eq:SNR_sum_TRX}
\end{equation}
where we consider:
\begin{equation}
    \text{SNR}_{\text{ASE}} = \frac{\text{ROP}}{P_{\text{ASE}}}\biggr|_{\textit{Bw}=R_s}
\end{equation} 
where ROP is the received optical power, and $P_{\text{ASE}}$ is the total power of the ASE noise injected in the optical link.
The only impairment in the optical domain (other than filtering, which will be taken into account later on) is ASE noise, because we neglect the fiber nonlinearity. Moreover, we take the symbol rate as a reference bandwidth to compute the noise power.
The transceiver's contribution is further decomposed into transmitter and receiver components:
\begin{equation}
\text{SNR}_{\text{TRX}}^{-1} = \text{SNR}_{\text{TX}}^{-1} + \text{SNR}_{\text{RX}}^{-1}
\end{equation}
The transmitter impact is considered constant due to fixed output power (typical values range from -10 dBm to 1 dBm), and it is assumed to have a negligible impact with respect to receiver's. The receiver contribution varies with the ROP, primarily due to photodetector noise sources including thermal, shot, and dark current noise.

To model transceiver performance, the effective transceiver $\text{SNR}_{\text{TRX}}$ is determined via a simplified curve fitting, based on experimental data and the equation and procedure proposed in \cite{Toru2024,TesiAndrea}. The formula, in linear scale, follows:
\begin{equation}
\text{SNR}_{\text{TRX}} = N\cdot \frac{\text{ROP}}{\text{ROP} + D}
\label{eq:trx_characterization}
\end{equation}
where ROP is expressed in mW and $N$ and $D$ are two fitting parameters to obtain the dashed line in \autoref{fig:transceiver}. To plot the back-to-back characterization, conversion to dB scale was performed on ROP and $\text{SNR}_{\text{TRX}}$ for better visualization.

As reported in \autoref{table:symbol_rates}, both transceivers operate at the same symbol rate for $\text{DP-QPSK-200G}$ and $\text{DP-16QAM-400G}$ modulation formats. Since the internal noise contributions of the device are independent of the modulation format, the fitting parameters $N$ and $D$ can be assumed to be the same for both cases and are therefore fitted simultaneously in \autoref{fig:transceiver}.
This property is particularly advantageous given the different characteristics of the two modulation formats. In particular, DP-QPSK is more robust and can therefore be used to characterize the device under low-ROP conditions. Conversely, DP-16QAM is more suitable at high ROP values, where DP-QPSK yields extremely low BER levels. In such conditions, BER measurements with DP-QPSK become inaccurate and tend to saturate between $\text{BER} = 10^{-10}$ and $\text{BER} = 10^{-9}$ since typically in commercial transceivers BER estimation is performed only on the overhead bits. The saturation is depicted with empty bullets in \autoref{fig:transceiver}.

The experimental setup to obtain the curves of \autoref{fig:transceiver} was prepared by excluding external noise sources in a back-to-back transceiver connection whose signal can be attenuated using a Variable Optical Attenuator (VOA).
By changing the attenuation we tested different ROP conditions, enabling the generation of \( \text{SNR}_\text{TRX}(\text{ROP}) \) curves by mapping BER measurements to SNR values inverting \eqref{eq:BER_vs_SNR} with the proper modulation format coefficients $k_{1,2}$, remembering that for this characterization, $\text{SNR}_\text{TRX}$ is the only relevant SNR contribution. 
The resulting curves demonstrate monotonic SNR growth towards saturation at high ROP.

\subsection{D-transform and power spectra evaluation for wide-sense stationary processes}

Since this work deals with equalizers operating in discrete domain, it is necessary to introduce an effective tool which allows to operate with discrete sequences and discrete filters, avoiding the complexity that a time-domain representation. Such tool is the D-transform\cite{cioffiD}, which is defined for sequences $x_k$ defined $\forall k \in (-\infty,\infty)$ with $k\in \mathbb{Z}$ as ($x_k$ can be complex):
\begin{equation}
    X(D)=\sum_{k=-\infty}^{\infty}x_k \cdot D^k \quad \quad \forall D \in \mathcal{D}_x, \quad D \in \mathbb{C}
\end{equation}
Where $\mathcal{D}_x$ is the region of convergence of complex D values, for which the sum $X(D)$ converges, and $X(D)$ is analytic in $\mathcal{D}_x$.
The inverse transform is a clockwise line integral around any closed circle in the region of convergence:
\begin{equation}
    x_k=\frac{1}{2 \pi j} \oint_{D \in \mathcal{D}_x} X(D) \cdot D^{k-1} dD
\end{equation}
For applications requiring Z-transform, it is sufficient to impose $Z=D^{-1}$. The sequence $x_{-k}^*$ has D-transform $X^*(D^{-*})=\sum_{k=-\infty}^{k=\infty}x_k^*D^{-k}$. An important property of D-transform is that, when the region of convergence includes the unit circle (which is true for all the sequences considered in this work), and the discrete sequence is obtained by sampling a continuous time signal at the symbol rate $T$, the discrete-time sequence's Fourier transform exists as: 
\begin{equation}
    X\left( e^{-j 2 \pi f T} \right) = X(D)|_{D=e^{-j 2 \pi f T}}
\end{equation}
The previous property is important when dealing with sequences representing wide-sense stationary processes, for which the evaluation of autocorrelation and power spectrum is often needed. Following \cite{cioffiD}, we recall that if $x_k$ is any stationary complex sequence, its autocorrelation function is $r_{xx,j} = \mathbb{E}[x_kx_{k-j}^*]$ with D-Transform:
\begin{equation}
    R_{xx}(D)=\mathbb{E}[X(D) \cdot X^*(D^{-*})] 
\end{equation}
The power spectrum of a stationary sequence is the Fourier transform of its autocorrelation function:
\begin{equation}
    R_{xx}\left(e^{j2\pi fT} \right) = R_{xx}(D)|_{D=e^{-j 2 \pi f T}} \quad -\frac{1}{2T} < f < \frac{1}{2T} 
\end{equation}
By stationarity $r_{xx,j}=r^*_{xx,-j}$ and $R_{xx}(D)=R_{xx}^*(D^{-*)})$, so the power spectrum is real and nonnegative $\forall f$. The random process energy can be computed as:
\begin{align}
    \notag
    \mathcal{E}_x &= \mathbb{E}[|x_k|^2] = T \cdot r_{xx,0} \\
    &=T \int_{-\frac{1}{2T}}^{\frac{1}{2T}}R_{xx}\left(e^{j2\pi f T}\right)df
\end{align}
If the sequence is deterministic then the power spectrum is the squared magnitude of its Fourier transform.

\begin{figure*}[htb!]
    \begin{center}
        \includegraphics[width=1\linewidth]{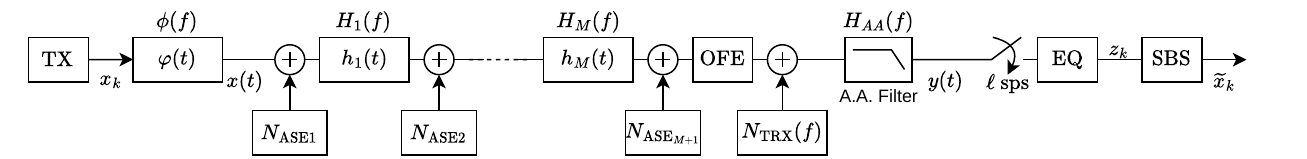}
        \caption{Reference optical link abstraction for the derivation of the equalization models.}
        \label{fig:block_diagram}
    \end{center}
\end{figure*}

\section{MODELS}
\label{sec:models}

In this section, we present a comprehensive framework for equalization models tailored to the specific requirements and characteristics of optical communication systems.

We consider both infinite- and finite-length equalizers and introduce a unified formalism for their implementation that encompasses the various abstractions proposed in the literature. Beyond reviewing these models, we generalize them to the case of arbitrary transceiver-noise PSD, with the goal of providing a rigorous and practically relevant tool for ONDT implementation.

For each model, we highlight its main strengths, limitations, and simplifying assumptions, in order to clarify when a given abstraction should be preferred and how the appropriate level of complexity can be selected according to the target application, since in many cases a simpler model may provide sufficient accuracy without requiring a more realistic and computationally demanding description.

\subsection{Optical Link Abstraction}
\label{sec:opt_link_abs}

The starting point for the analytical determination of the filtering penalty in optical systems is the block diagram shown in \autoref{fig:block_diagram}. A general optical link has been theoretically replicated. We consider dual-polarization coherent transmission and focus on evaluating the SNR for the symbol sequence transmitted through one of the two polarizations. This approach is justified, since the filtering effect acts identically on both polarizations. Therefore, when computing the total SNR as the average of the SNR on the two polarizations, we obtain the same result. In addition, we assume the modulation uses In-phase and Quadrature components, as is done in modern optical transceivers.

The transmitted symbols are $\{ x_k\}_{k=0}^{K-1}$, for $K$ successive transmissions, for each polarization, and they are generated at symbol rate equal to $R_s$ (from now on, we will refer also to the symbol period as: $T=\frac{1}{R_s}$), therefore we can write the sequence energy.
\begin{equation}
\mathcal{E}_x=\mathbb{E}\left[|x_k|^2\right]
\end{equation}

Consequently, the total transmitted signal power, accounting for both polarizations, is computed as follows:
\begin{equation}
    P_{\text{TX}}=2 \cdot \frac{\mathcal{E}_x}{T} = 2 \cdot \mathcal{E}_x \cdot R_s
\end{equation}

For derivation purposes, it is useful to write also the per-dimension energy, namely the energy of the In-phase or Quadrature component of one of the two polarizations:
\begin{equation}
    \bar{\mathcal{E}}_x=\frac{1}{2}\mathcal{E}_x=\frac{P_{\text{TX}}}{4\cdot R_s}
\end{equation}

The $x_k$ symbols are modulated by the shaping filter $\varphi(t)$ (assumed to satisfy the Nyquist criterion for intersymbol interference (ISI) free channels and symbol-by-symbol (SBS) detection, and with Fourier transform \mbox{$\Phi(f)=\mathcal{F}[\varphi(t)]$}), so we can write the modulator output.
\begin{equation}
    x(t)=\sum_{k=0}^{K-1}x_k \cdot \varphi(t-kT)
\end{equation}

The modulated symbols pass through a cascade of $M$ optical filters described by their transfer functions $H_i(f)$ ($i=1,\dots,M$) experiencing the injection of distributed noise along the line, before the AA filter $H_{AA}(f)$. The signal is then sampled at $\ell$ samples per symbol (in commercial transceivers typically $\ell = 2$). After sampling, the signal is equalized, and SBS detection is performed on the $z_k$ ($k=0,\dots , K-1$) outputs of the equalizer, providing the final estimates $\hat{x}_k$ of the input symbols. 

This model incorporates some simplifying assumptions. As anticipated in the previous section, we can assume the absence of the NLI, because metro and access networks are characterized by short links and low optical power \cite{roadmap}. Nevertheless, as proposed in \cite{RossoICTON:2025}, NLI can be conservatively accounted for if included as a Gaussian noise source concentrated on the receiver side. 

All the ASE noise sources are assumed to be characterized by a flat PSD with a value of $N_\text{ASE,i}$ (where $i = 1, ..., M+1$) for each ASE source, at least within the signal bandwidth. Therefore, each ASE noise contribution is an additive white Gaussian noise when injected into the link. The PSD of ASE noise due to optical amplification can be computed using the well-known formula\cite{agrawal}.
\begin{equation}
\label{eq:ASE_noise}
    N_{\text{ASE}_i}= \frac{1}{4}\sum_{j=1}^{L_{i}}hf_0(G_{j,i}-1)\textit{NF}_{j,i}
\end{equation}
where $L_{i}$ is the number of amplified span located in-between each pair of optical filters, $h$ is the Planck constant, $f_0$ is the central frequency of the transmitted signal, $G_{j,i}$ and $\textit{NF}_{j,i}$ are the amplifier gain and noise figure of each amplifier. The factor $\frac{1}{4}$ is needed because we consider the noise PSD for each modulation component (In-phase and Quadrature) and for each polarization.

According to the transceiver model introduced in the previous Section, and represented in \autoref{fig:trx_block_diagram}, we consider the transceiver noise to incorporate the OFE impairments and to be concentrated at the receiver side, with a PSD assuming a generic shape $N_\text{TRX}(f)$. 
\begin{equation}
    \label{eq:trx_psd} 
    N_\text{TRX}(f) = \dot{N}_\text{TRX} \cdot S_\text{TRX}(f)
\end{equation}
where $\dot{N}_\text{TRX}$ is the white-noise equivalent PSD value, expressed in mW/Hz, and $S_\text{TRX}(f)$ is the shape of the colored PSD (adimensional). They depend on $\text{SNR}_\text{TRX}$ and on the specific commercial transceiver under analysis. If the PSD of the transceiver noise is assumed to be white, one can simply set $S_\text{TRX}(f)=1$ and compute $\dot{N}_\text{TRX}$ as follows.
\begin{equation}
    \label{eq:white_trx_psd}
    \dot{N}_\text{TRX}=\frac{\text{ROP}}{R_S\cdot \text{SNR}_\text{TRX}}
\end{equation}

\subsection{Noise Whitening}
\label{sec:noise_whitening}

\begin{figure}[htb!]
    \centering
    \includegraphics[width=1\linewidth]{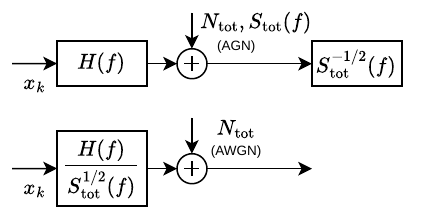}
    \caption{Noise whitening scheme. AGN = Additive Gaussian Noise; AWGN = Additive White Gaussian Noise.}
    \label{fig:whitening_filter}
\end{figure}

 In \autoref{fig:block_diagram} filtering stages contribute to make the total PSD of the noise at the receiver side to be colored. In such a case, noise whitening approach is widely used in digital communications to simplify models.
 Exploiting the white-noise equivalent channel representation it is possible to derive without loss of information the models of ZFE, MMSEE and FSE Equalizers with a uniform and elegant mathematical framework. In the following Sections, we will also consider the WFLE, based on noise whitening, which is a reliable approximation already successfully exploited in past works \cite{miotto2025arxive,sdm_finite_taps}.
 
 In \autoref{fig:whitening_filter} it is illustrated how to obtain a white-noise equivalent channel model, given the condition that the colored noise PSD has an invertible square root\cite{Proakis2008, cioffi}. In this section we derive the quantities $H(f)$, $S_\text{tot}(f)$ and $N_\text{tot}$, which are the white-noise equivalent channel frequency response, the normalized colored noise PSD, and the white equivalent PSD, respectively.
 The initial step involves evaluating the noise on the receiver side, which is the sum of multiple components colored on the basis of the number and shape of crossed filters.
 Using the well-known formula to calculate the output PSD of a random process crossing a filter, knowing the input PSD we derive the normalized colored PSD before sampling:
\begin{align}    
    \label{eq:PSD_colored}
    \notag
    \textit{S}_{\text{tot}}(f) &= \frac{ \sum_{i=1}^{M+1}N_{ASE,i}  \prod_{n=i}^{M}|H_n(f)|^2|H_{AA}(f)|^2}{N_{\text{tot}}} + \\
    & +  \frac{ (\dot{N}_\text{TRX} S_\text{TRX}(f))|H_{AA}(f)|^2}{N_{\text{tot}}}
\end{align}
where $\prod_{n=M+1}^M|H_n(f)|^2=1$,  the PSD of the transceiver noise is taken from \eqref{eq:trx_psd}, and the total colored noise PSD has been normalized by the following factor.
\begin{equation}
    \label{eq:total_sigma}
    N_{\text{tot}}=\sum_{i=1}^{M+1}N_{ASE,i} + \dot{N}_\text{TRX}
\end{equation}
Since the penalty on the link is due to linear filtering, the colored PSD expressed by \eqref{eq:PSD_colored} has an invertible square root. Moreover, due to the normalization factor, the  PSD of the equivalent white noise source on the receiver side is $N_{\text{tot}}$.

Following the notation of \cite{cioffi}, we introduce the white-noise equivalent pulse response of the optical link as the inverse Fourier transform of the white-noise equivalent channel frequency response.
\begin{align}
    \label{eq:h_pulse}
   h(t) &= \mathcal{F}^{-1}[H(f)]\\
   \label{eq:wn_eq_ch_func}
   H(f) &= \frac{\Phi(f)\cdot \prod_{i=1}^M|H_i(f)||H_{AA}(f)|}{\sqrt{S_{\text{tot}}(f)}}
\end{align}
where the numerator in \eqref{eq:wn_eq_ch_func} has been obtained from the cascade of the shaping filter optical filters, and AA filter, with $\Phi(f)$ being the Fourier transform of the pulse response of the shaping filter. The denominator is due to the effect of the whitening filter according to \autoref{fig:whitening_filter}. Notably, the AA filter appears both in the numerator and in the denominator, therefore cancels out, and does not contribute to the white-noise equivalent pulse response.

Since the pulse energy is not necessarily normalized to 1, it is useful to introduce the normalized pulse response as follows.
\begin{equation}
    \varphi_h(t) \triangleq \frac{h(t)}{||h||}
\end{equation}
where
\begin{equation}
    ||h|| = \sqrt{\int_{-\infty}^{\infty}h(t)\cdot h^*(t)dt}=\sqrt{\langle h(t), h(t) \rangle}
\end{equation}
so that the band limited channel output before noise injection is as follows.
\begin{equation}
    x_h(t) = \sum_{k=0}^{K-1}x_k\cdot||h||\cdot \varphi_h(t-kT)
\end{equation}

\subsection{THE UNFILTERED SNR BOUND}
\label{sec:MFB}

In the optical system under study, the passband effect of the optical filters limits the channel, causing ISI that affects the performance in terms of SNR degradation. The maximum SNR is obtained when the filtering effect is negligible and therefore when the signal is not distorted. Such a situation can occur when the band of the filters is large compared to the signal bandwidth. When this happens, the SNR computation is straightforward, because ASE noise is flat in the signal bandwidth. We conclude that in this case the SNR computed taking into account ASE noise and transceiver noise, neglecting NLI, is as follows.
\begin{equation}
    \text{SNR}_\text{max} = \left( \text{SNR}_{\text{ASE}}^{-1} + \text{SNR}_{\text{TRX}}^{-1} \right)^{-1} = \text{SNR}
\end{equation}
where $\text{SNR}_{\text{TRX}}$ is given in the Transceiver Section. Since the signal and noise power can be computed, we can explicitly express the SNR bound according to the notation developed for noise whitening in Section \ref{sec:models}-\ref{sec:noise_whitening}, which will be useful when deriving most of the considered equalization models, allowing to compare their performance.
\begin{equation}
    \text{SNR}=\frac{\text{ROP}}{P_{\text{ASE}}+P_{\text{TRX}}}= \frac{4\cdot \bar{\mathcal{E}}_x\cdot R_s \cdot ||h||^2}{4 \cdot N_{\text{tot}} \cdot R_s}
 =\frac{\bar{\mathcal{E}}_x \cdot ||h||^2}{N_{\text{tot}}}
\end{equation}
where we point out that bandwidth terms cancel out since both signal and noise power are computed with respect to the same band equal to the symbol rate.

\subsection{INFINITE LENGTH EQUALIZERS}
\label{sec:infinite_equalizers}

\begin{figure}[htbp]
    \begin{center}
    \includegraphics[width=1\linewidth]{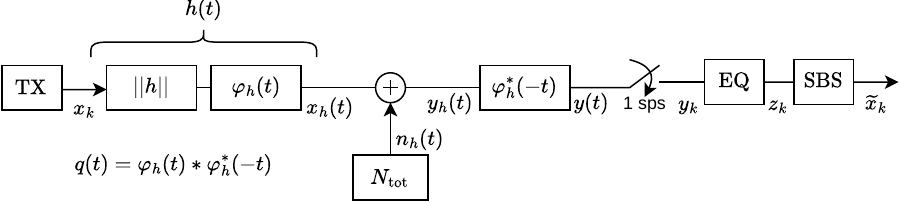}        
    \end{center}
    \caption{Zero Forcing and Minimum Mean Square Error Equalizers reference block diagram for whitened noise.}
    \label{fig:channel_model}
\end{figure}

In this section we examine the infinite length equalization models, namely the ZFE, the MMSEE and the FSE. Regarding ZFE and MMSEE, some additional assumptions and definitions are needed, since these models are based on matched filter theory. The optical link in \autoref{fig:channel_model} represents the white-noise equivalent link for ZFE and MMSEE equalization. After matched filtering, the DSP is assumed to perform equalization operating at a sampling rate equal to 1 sample per symbol. The matched filtering assumption is not realistic in optical systems. In general, the presence of a matched filter before sampling is not necessary, since modern commercial transceivers implement both matched filtering and equalization in the discrete domain. We will get rid of matched filtering when considering the FSE, WFLE, and FLE models. 


We define the deterministic autocorrelation function $q(t)$ as follows.
\begin{equation}
    q(t) \triangleq \varphi_h(t)*\varphi_h^*(-t)=\frac{h(t)*h^*(-t)}{||h||^2}
\end{equation}
where $*$ is the complex conjugate operation. It is useful to note that $q(t)$ is Hermitian \mbox{($q(t)=q^*(-t)$)} and that $q(0)=1$.

Since in the considered optical system we assume that we know the expression of each filter and each noise contribution, it is possible to rewrite $Q(f)=\mathcal{F}[q(t)]$ explicitly.
\begin{align}
    \label{eq:Q_explicit}
    Q(f) &= \frac{|H(f)|^2}{||h||^2} = \frac{\left\vert \Phi(f)\cdot \prod_{i=1}^{M}H_i(f) \cdot H_\text{AA}(f) \right\vert^2}{||h||^2 \cdot S_{\text{tot}}(f)}
\end{align}

Once the channel model has been established, it is possible to derive the filtering penalty model according to the chosen equalization strategy based on the steps followed in \cite{cioffi}. In the following, we provide the expression of the SNR after equalization and the filtering penalty coefficient for all the considered equalization strategies.

\subsubsection{Zero Forcing Equalizer}
\label{sec:ZFE}

The reference system for the current model is provided in \autoref{fig:channel_model}. ZFE is the simplest and most intuitive equalizer among the ones considered in this work. However, it is also the worst in terms of performance, due to its noise enhancement effect. Nevertheless, its model allows one to express the filtering penalty in a disaggregated manner, based on a cumulative metric\cite{gnpy}, as will be clear in the following and as shown in \cite{ONDM2025}. This feature is pivotal when including the filtering penalty in a ONDT transmission model, therefore, ZFE is part of the present discussion. Moreover, in many practical applications in which filtering and noise are limited, the performance assessment given by the ZFE is very accurate yet conservative, as quantitatively assessed in the simulations in the following.
Based on the derivation provided in \cite{cioffi}, the output SNR of the ZFE, expressed as a degradation of the unfiltered SNR is as follows.
\begin{equation}
    \label{eq:SNR_ZFE}    \text{SNR}_{\text{ZFE}}=\frac{\text{SNR}}{k_{\text{ZFE}}}
\end{equation} 

The filtering penalty factor $k_\text{ZFE}$ is obtained based on the following expression.
\begin{equation}
    \label{eq:kZFE}
    k_{\text{ZFE}}=\frac{1}{R_s}\int_{-\frac{R_s}{2}}^{\frac{R_s}{2}}\frac{1}{Q\left(e^{-j\frac{2\pi f}{R_s}}\right)}df
\end{equation}
where, compared to \cite{cioffi}, we substituted the symbol period with the inverse of the symbol rate $T=\frac{1}{R_s}$, we used the frequency instead of angular velocity $w=2\pi f$ and we substituted the noise PSD with the one previously computed in \eqref{eq:total_sigma}, namely: $\frac{\mathcal{N}_0}{2}=N_{\text{tot}}$. The term $Q\left(e^{-j\frac{2\pi f}{R_s}}\right)$ is the folded spectrum of $Q(f)$ expressed in \eqref{eq:Q_explicit}, it is periodic with period $R_s$ and it can be written as:
\begin{equation}
    Q\left(e^{-j\frac{2\pi f}{R_s}}\right) = R_s \cdot \sum_{n=-\infty}^{\infty} Q\left( f+ nR_s \right)
\end{equation}
For practical implementations, one can compute $Q\left(e^{-j\frac{2\pi f}{R_s}}\right)$ based on \eqref{eq:Q_explicit} performing a periodic repetition in frequency or numerically by undersampling in time-domain.

The disaggregation of ZFE can be performed by considering each noise source characterized by $N_{ASE,i}$ \mbox{($i = 1,\dots,M+1$)} and $N_\text{TRX}(f)$ individually and computing the corresponding constant $k_i$ through \eqref{eq:kZFE}, with the other sources injecting no noise in order to evaluate the colored PSD of \eqref{eq:PSD_colored}. The final formula for computing $k_i$ follows.

\begin{align}
    \label{eq:disaggregated_model}
    k_{ASE,i} &= \frac{1}{R_s} \bigintss_{-\frac{R_s}{2}}^{\frac{R_s}{2}} \frac{ \prod_{n=i}^{M}\left| H_n\left(e^{j\frac{2\pi f}{R_s}}\right) \right|^2}{\left|\Phi\left(e^{j\frac{2\pi f}{R_s}}\right) \right|^2 \prod_{m=1}^{M}\left| H_m\left(e^{j\frac{2\pi f}{R_s}}\right)\right|^2}df \\
    k_\text{TRX} &= \frac{1}{R_s} \bigintss_{-\frac{R_s}{2}}^{\frac{R_s}{2}} \frac{ S_\text{TRX}\left(e^{j\frac{2\pi f}{R_s}}\right)}{\left|\Phi\left(e^{j\frac{2\pi f}{R_s}}\right) \right|^2 \prod_{m=1}^{M}\left| H_m\left(e^{j\frac{2\pi f}{R_s}}\right)\right|^2}df
\end{align}

This closed-form disaggregated formulation is especially valuable for ONDTs, as it allows filtering penalties to be embedded directly into the transmission model in an interpretable and scalable way, preserving the contribution of the individual network elements and supporting impairment-aware performance prediction.

\subsubsection{Minimum Mean Square Error Equalizer}

The MMSEE equalizer balances ISI reduction and noise enhancement. In this case, the reference system is the same as ZFE, reported in \autoref{fig:channel_model}. 
As mentioned in Section \ref{sec:models}-\ref{sec:infinite_equalizers}-\ref{sec:ZFE} it is possible to define a penalty due to filtering $k_\text{MMSEE}$ with respect to the SNR at unfiltered bound.
\begin{equation}
    \label{eq:SNR_MMSEE}
    \text{SNR}_{\text{MMSEE}} = \frac{\text{SNR}}{k_\text{MMSEE}}-1
\end{equation}
where the term “$-1$” is needed to obtain the unbiased SNR, since the MMSEE equalizer is by definition biased \cite{cioffi}. The explicit expression of $k_\text{MMSEE}$ is derived in \cite{cioffi}, here we report the final result.
\begin{equation}
    \label{eq:k_MMSEE}
    k_\text{MMSEE} = \frac{1}{R_s} \int_{-\frac{R_s}{2}}^{\frac{R_s}{2}}\frac{1}{\left( Q\left(e^{-j\frac{2 \pi f}{R_s}}\right)+\frac{1}{\text{SNR}}\right)}df
\end{equation}
where $R_s$ is the symbol rate, $Q\left(e^{-j\frac{2 \pi f}{R_s}}\right)$ is computed as in the ZFE case, and the additional positive term $\frac{1}{\text{SNR}}$, depending on the unfiltered bound for the SNR, is the only difference compared to the ZFE. This term represents the ability of the MMSEE equalizer to balance ISI reduction and noise enhancement. 

\subsubsection{Fractionally Spaced Equalizer}

\begin{figure}[htbp]
    \begin{center}
        \includegraphics[width=1\linewidth]{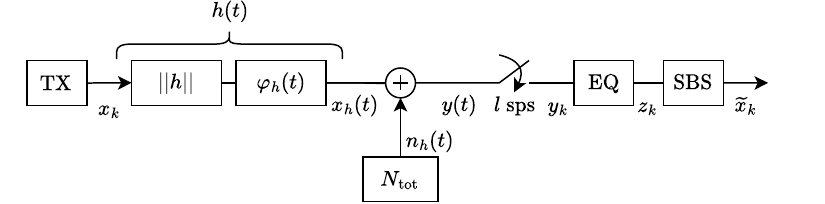}
        \caption{Fractionally Spaced  and Finite Length Equalization reference block diagram for whitened noise.}
        \label{fig:FSE}
    \end{center}
\end{figure}

In this section, the reference channel model block scheme is provided in \autoref{fig:FSE}. The FSE exploits the minimum mean-square error equalization strategy, with some additional features that allow one to increase the performance by a slight increment in terms of complexity.

In this more realistic model, we eliminate the matched filter before sampling. The DSP performs both equalization and matched filtering in the discrete domain, according to what happens in commercial transceivers deployed in optical systems. In addition, the FSE increases the sampling rate by some rational number $\ell$, (\mbox{$\ell>1$}), which we assume to be an integer. The unbiased SNR for the FSE is derived in \cite{cioffi}, and the final expression is as follows.
\begin{equation}
    \label{eq:SNR_FSE}
    \text{SNR}_{\text{MMSEE,FSE}} = \frac{\text{SNR}}{k_\text{MMSEE,FSE}}-1
\end{equation}

The corresponding filtering penalty follows.
\begin{equation} 
    \label{eq:k_FSE}
    k_\text{MMSEE,FSE}=\frac{1}{R_s} \int_{-\frac{R_s}{2}}^{\frac{R_s}{2}}\frac{\ell}{\left| \left| \mathbf{H}\left(e^{-j\frac{2\pi f}{R_s}}\right) \right| \right|^2+\frac{\ell}{\text{SNR}}}df
\end{equation}
where the parameter $\left| \left| \mathbf{H}\left( e^{-j\frac{2\pi f}{R_s}} \right) \right| \right|^2$ accounts for the shape of the filters and for the PSD.
\begin{equation}
    \left| \left| \mathbf{H}\left( e^{-j\frac{2\pi f}{R_s}} \right) \right| \right|^2=\sum_{i=1}^{\ell} \left| H_i\left( e^{-j\frac{2\pi f}{R_s}}\right) \right|^2
\end{equation}
and it can be computed from the symbol-rate-spaced phases $H_i(D)$, evaluated in the unit circle. The expression of $H_i\left( e^{-j\frac{2\pi f}{R_s}} \right)$ is:
\begin{equation}
    \label{eq:Hi}
    H_i\left( e^{-j\frac{2\pi f}{R_s}} \right)=\sum_{k=-\infty}^{\infty}h\left(kT-(i-1)\frac{T}{\ell}\right) \cdot D^k \bigg|_{D=e^{-j\frac{2\pi f}{R_s}} }
\end{equation}

For practical implementation purposes, we recall that the white-noise equivalent pulse response of the optical link $h(t)$ is defined in \eqref{eq:h_pulse}. Therefore, one can sample $h(t)$ to obtain the $\ell$ phases and then perform a Fourier Transform to evaluate $H_i\left( e^{-j\frac{2\pi f}{R_s}} \right)$.

The structure of \eqref{eq:k_FSE} is similar to the filtering-penalty coefficient of the MMSEE equalizer in \eqref{eq:k_MMSEE}. Indeed, the FSE is also based on the minimum mean-square error criterion and therefore can mitigate ISI while limiting noise enhancement. Moreover, practical adaptive implementations of the FSE based on LMS or constant modulus algorithms have been shown to converge to MMSEE-equalizer performance \cite{CMA_vs_MMSEE}. In the present analysis, timing-related effects are not considered; therefore, the possible additional benefit of the FSE observed in practical implementations, in particular its improved robustness to sampling-phase errors\cite{Proakis2008, cioffi}, is beyond the scope of this discussion.

\subsection{FINITE LENGTH EQUALIZERS}
\label{sec:FLE}

The models analyzed previously assume infinite-length filtering. In practical applications, the equalization filter is implemented as a finite-impulse-response (FIR) filter. This is because adaptive equalization techniques strongly exploit FIR structures. The models proposed in this section are based on MMSEE equalization.

\subsubsection{Whitened Finite Length Equalizer}
\label{sec:FLE_whitening}

As done for the FSE case, the reference block scheme is depicted in \autoref{fig:FSE}. At this stage, we keep the previously defined channel function $h(t)$ embedding the PSD of the colored noise, therefore, assuming whitening filtering, as done in \cite{cioffi, sdm_finite_taps, miotto2025arxive}.

We recall that $\ell > 1$ is the oversampling factor. Therefore, after sampling at $t=kT-\frac{iT}{\ell}$ ($i=0,\dots,\ell-1$) the channel output becomes as follows.
\begin{align}
    \notag
    y\left(kT-\frac{iT}{\ell}\right) = \sum_{m=-\infty}^{\infty}x_m & \cdot h\left( kT-\frac{iT}{\ell}-mT \right)+ \\
    &+ n\left( kT-\frac{iT}{\ell} \right) 
\end{align}
where the variance of the sampled noise, as for the FSE, is equal to $\ell \cdot N_\text{tot}$. 

It is possible and convenient to represent the channel using vectors and matrices, by grouping the $\ell$ phases per symbol period of the oversampled $y(t)$.
\begin{equation}
    \mathbf{y}_k=\sum_{m=-\infty}^{\infty}x_{k-m} \cdot \mathbf{h}_m + \mathbf{n}_k
\end{equation}
where:
\begin{align}
    \notag \mathbf{y}_k&=\begin{bmatrix}
        y(kT), & y(kT-\frac{T}{\ell}), & \dots & y(kT-\frac{\ell-1}{\ell}T)
    \end{bmatrix}^T \\
    \notag \mathbf{h}_k&=\begin{bmatrix}
        h(kT), & h(kT-\frac{T}{\ell}), & \dots & h\left(kT-\frac{\ell-1}{\ell}T\right)
    \end{bmatrix}^T \\
    \mathbf{n}_k&=\begin{bmatrix}
        n(kT), & n(kT-\frac{T}{\ell}), & \dots & n\left(kT-\frac{\ell-1}{\ell}T\right)
    \end{bmatrix}^T
\end{align}

The analysis in this section is carried in time-domain, based on some assumptions needed to describe the discretized channel in matrix form. The step-by-step derivation is provided in \cite{cioffi}. Here we report the main passages and considerations that are useful to understand the final expression.

The channel pulse response $h(t)$ is assumed to extend over a finite time interval, so any nonzero value of $h(t)$ is considered negligible outside of this interval, which is defined as $-(\ell -1)T/\ell \leq t \leq \nu T$. Therefore, \mbox{$\mathbf{h}_k \approx 0$} for $k<0$ and $k>\nu$. If this does not happen, the model accuracy is reduced due to an arbitrary truncation of the channel. In most cases, it is possible to find the value of $\nu$ to avoid this approximation despite the fact that the whitening filter tends to create long temporal tails in $h(t)$. Further discussion on the limit of this model is provided in the next section, which focuses on the FLE mathematical model without the whitening filtering assumption. Moreover, a performance comparison is carried out in the next Section.

We consider that $N_f$ successive $\ell$-tuples of samples of $y(t)$ are exploited to perform equalization. The coefficient $N_f$ is related to the number of taps of the FLE, which can be calculated from the product $\ell \cdot N_f$. The final assumption is that $\ell$ is an integer, which leads to a simpler expression for the channel matrix. Based on the previous assumptions, the channel model is as follows.
\begin{align}
    \notag \mathbf{Y}_k &= \begin{bmatrix}
        \mathbf{y}_k, & \mathbf{y}_{k-1}, & \dots & \mathbf{y}_{k-N_f+1}
    \end{bmatrix}^T = \\
    & \notag = \begin{bmatrix}
        \mathbf{h}_0 & \mathbf{h}_1 & \dots & \mathbf{h}_\nu & 0 & 0 & \dots & 0 \\
        0 & \mathbf{h}_0 & \mathbf{h}_1 & \dots & \mathbf{h}_\nu & 0 & \dots & 0 \\
        \vdots & \vdots & \ddots & \ddots & \ddots & \ddots & \vdots & \\
        0 & \dots & 0 & 0 &\mathbf{h}_0 & \mathbf{h}_1 & \dots & \mathbf{h}_\nu
    \end{bmatrix} \cdot \\
    \label{eq:bigchannelmatrix} & \cdot \begin{bmatrix}
        x_k \\ x_{k-1} \\ \vdots \\ \vdots \\ x_{k-N_f-\nu +1}
    \end{bmatrix} +
    \begin{bmatrix}
        \mathbf{n}_k \\
        \mathbf{n}_{k-1} \\
        \vdots \\
        \mathbf{n}_{k-N_f+1}
    \end{bmatrix}
\end{align}

The $(N_f\cdot \ell)\times(N_f+\nu)$ matrix in \eqref{eq:bigchannelmatrix} is denoted as $\mathbf{H}$, while the data vector and the noise vector are defined, respectively, as $\mathbf{X}_k$ and $\mathbf{N}_k$. Then the compact form of \eqref{eq:bigchannelmatrix} becomes:
\begin{equation}
    \label{eq:matrix_channel}
    \mathbf{Y}_k=\mathbf{H} \cdot \mathbf{X}_k + \mathbf{N}_k
\end{equation}
In addition to the noise variance information, $\mathbf{H}$ fully describes the filtering penalty for finite-length equalization.

The WFLE unbiased SNR, with whitening filtering, is obtained as follows.
\begin{equation}
    \label{eq:SNR_FLE}    \text{SNR}_\text{MMSEE,WFLE}=\frac{\bar{\mathcal{E}}_x}{\sigma^2_\text{MMSEE,WFLE}}-1
\end{equation}
where the post-equalization noise variance is:
\begin{equation}
    \label{eq:variance_FLE}
    \sigma^2_\text{MMSEE,WFLE} = \bar{\mathcal{E}}_{\mathbf{x}} - \mathbf{w} \cdot R_{\mathbf{Y}\mathbf{Y}} \cdot \mathbf{w}^\dag
\end{equation}
The symbol $\dag$ stands for the transpose conjugation operator.

The expression of the noise variance after the equalizer in \eqref{eq:variance_FLE} depends on the vector of the equalizer coefficients $\mathbf{w}$.
\begin{equation}
    \mathbf{w}= R_{x\mathbf{Y}}\cdot R_{\mathbf{YY}}^{-1}
\end{equation}

To express $\sigma^2_\text{MMSEE-WFLE}$ and $\mathbf{w}$, it is necessary to compute the correlation matrices $R_{\mathbf{YY}}$ and $R_{x \mathbf{Y}}$, which are the autocorrelation matrix of the output samples and the cross-correlation matrix of the input-output samples, respectively. Both $R_{\mathbf{YY}}$ and $R_{x \mathbf{Y}}$ depend on the channel matrix $\mathbf{H}$ and the noise autocorrelation matrix $R_{\mathbf{NN}}=\ell \cdot N_\text{tot} \cdot \mathbf{I}$. The noise autocorrelation matrix is diagonal since we consider whitened noise.
\begin{align}
    \label{eq:output_corr}
    R_{\mathbf{YY}} &= \bar{\mathcal{E}}_{\mathbf{x}} \cdot \mathbf{H} \mathbf{H}^\dag + R_{\mathbf{NN}} \\
    \label{eq:inout_corr}
    R_{x \mathbf{Y}} &= \bar{\mathcal{E}}_{\mathbf{x}} \cdot [0\dots 0, \mathbf{h}_\nu ^\dag \dots\mathbf{h}_0 ^\dag, 0\dots 0]
\end{align}

\subsubsection{Finite Length Equalizer}
\label{sec:FLE_colored}

A more realistic FLE model can be obtained by considering the initial link abstraction, depicted in \autoref{fig:block_diagram}, whose main features and assumptions were discussed in Section \ref{sec:models}-\ref{sec:opt_link_abs}. In this case, noise whitening is not applied before sampling. Then matched filtering and equalization are performed in discrete domain after oversampling by a factor $\ell$.

As anticipated, the ideal noise whitening leads to an approximation of the WFLE performance. The continuous-time whitening filter $1/\sqrt{S_\text{tot}}$ in \eqref{eq:h_pulse} can have infinite pulse response. This creates long tails in the equivalent channel pulse response $h(t)$. In such cases, the assumption $h(t) \approx 0$ for $t\leq -(\ell-1)T/\ell$ and $t \geq \nu T$ is not valid, causing a truncation of $h(t)$ when building the channel matrix $\textbf{H}$ defined in \eqref{eq:bigchannelmatrix} and \eqref{eq:matrix_channel}. 
In addition, if no whitening is applied to the noise before sampling, the noise autocorrelation matrix $R_{\mathbf{NN}}$ in \eqref{eq:output_corr} is not diagonal and must be computed based on the pulse responses experienced by each noise source in the link.

We define the channel pulse response as follows.
\begin{align}
    \label{eq:signal_pulse_resp}
    \tilde{h}(t) = \mathcal{F}^{-1}\left[ \Phi(f)\cdot \prod_{i=1}^M|H_i(f)|\cdot |H_\text{AA}(f)| \right]
\end{align}
Notably, the colored PSD is not included in the channel pulse response expression. Each noise source is a continuous time signal which is sampled at $\ell/R_s$ rate after the convolution with the filters. The channel pulse response for each ASE noise source is expressed as follows. 
\begin{align}
    \label{eq:noise_pulse_resp}
    \tilde{g}_i(t) = \mathcal{F}^{-1}\left[ \prod_{n=i}^{M} |H_n(f)| \cdot |H_\text{AA}(f)| \right] \quad i=1,\dots,M+1
\end{align}
where, as in \eqref{eq:PSD_colored}, we assume $\prod_{n=M+1}^M|H_n(f)|=1$.

Regarding the transceiver noise, it is necessary to consider the colored nature of the PSD that we are assuming, in addition to the AA filter.
\begin{equation}
    \label{eq:trx_pulse_resp}
    \tilde{g}_\text{TRX}(t) = \mathcal{F}^{-1}\left[ \sqrt{S_\text{TRX}(f)} \cdot |H_\text{AA}(f)| \right]
\end{equation}

As done in Section \ref{sec:models}-\ref{sec:FLE}-\ref{sec:FLE_whitening} for the WFLE, it is convenient to group the $\ell$ phases per symbol period of the pulse response of the channel as follows.
\begin{align}
    \mathbf{\tilde{h}}_k &= \begin{bmatrix}
        \tilde{h}(kT) & \tilde{h}(kT-\frac{T}{\ell}) & \dots & \tilde{h}(kT-\frac{(\ell - 1)}{\ell}T)
    \end{bmatrix}^T
    \label{eq:phases_colored}
\end{align}

To build finite-dimension matrices we also consider in this case that we collect the samples of $\tilde{h}(t)$ and $\tilde{g}_i(t)$ according to the factor $\nu$ in the time interval $-(\ell - 1)T/\ell \leq t \leq \nu T$. As stated at the beginning of this section, the absence of the whitening filter leads to higher accuracy than the WFLE. Without the whitening filter, the truncation of the time interval is expected to be much more accurate due to the fact that $\tilde{h}(t)$ and $\tilde{g}_i(t)$ are expected to be finite pulse responses.

Based on the pulse response vectors defined in \eqref{eq:phases_colored} and the pulse responses in \eqref{eq:noise_pulse_resp}, it is possible to rewrite the autocorrelation matrix of the output samples and the cross-correlation matrix of the input-output samples of \eqref{eq:output_corr} and \eqref{eq:inout_corr} as follows\cite{rizzelli_finitelength}.
\begin{align}
    \label{eq:colored_out_autocorr}
    \tilde{R}_{\mathbf{YY}} &= \bar{\mathcal{E}}_{\mathbf{x}} \mathbf{\widetilde{H}}  \mathbf{\widetilde{H}}^\dag + \sum_{i=1}^{M+1} N_{ASE,i} \mathbf{\widetilde{G}}_i  \mathbf{\widetilde{G}}^\dag_i + \dot{N}_\text{TRX} \mathbf{\widetilde{G}_\text{TRX}} \mathbf{\widetilde{G}_\text{TRX}}^\dag \\
    \label{eq:colored_inout_corr}
    \tilde{R}_{x \mathbf{Y}} &= \bar{\mathcal{E}}_{\mathbf{x}} \cdot [0\dots 0, \mathbf{\tilde{h}}_\nu ^\dag \dots\mathbf{\tilde{h}}_0 ^\dag, 0\dots 0]
\end{align}
In \eqref{eq:colored_out_autocorr} $\mathbf{\widetilde{H}}$ shares structural similarities with the channel matrix $\mathbf{H}$ defined in \eqref{eq:bigchannelmatrix} and \eqref{eq:matrix_channel}, being a $(N_f\cdot \ell)\times(N_f+\nu)$ Toeplitz matrix, which can be built from $\tilde{h}(t)$ as follows.
\begin{align}
    \mathbf{\widetilde{H}}=\begin{bmatrix}
            \tilde{\mathbf{h}}_0 & \tilde{\mathbf{h}}_1 & \dots & \tilde{\mathbf{h}}_\nu & 0 & 0 & \dots & 0 \\
        0 & \tilde{\mathbf{h}}_0 & \tilde{\mathbf{h}}_1 & \dots & \tilde{\mathbf{h}}_\nu & 0 & \dots & 0 \\
        \vdots & \vdots & \ddots & \ddots & \ddots & \ddots & \vdots & \\
        0 & \dots & 0 & 0 & \tilde{\mathbf{h}}_0 & \tilde{\mathbf{h}}_1 & \dots & \tilde{\mathbf{h}}_\nu
    \end{bmatrix}
\end{align}
In \eqref{eq:colored_out_autocorr} we also defined the matrices $\mathbf{\widetilde{G}}_i $ for $i=1,\dots,M+1$, which represent the channel matrix for each ASE noise source and are built starting from the pulse responses defined in \eqref{eq:noise_pulse_resp} as follows.
\begin{equation}
\notag
\dot{g}_{i,k,j} \triangleq \tilde{g}_i \left(kT-j\frac{T}{\ell} \right), \quad k = 0,\dots, \nu ; \quad j=0,\dots,\ell-1
\end{equation}
{\setlength{\arraycolsep}{2pt}
\begin{equation}
\label{eq:colored_noise_matrix}
\footnotesize
\mathbf{\widetilde{G}}_i =
\begin{bmatrix}  
\dot{g}_{i,0,\ell-1} & \dot{g}_{i,0,\ell-2} & \cdots & \dot{g}_{i,\nu,0} & 0 & 0 & \cdots & 0 \\
0 & \dot{g}_{i,0,\ell-1} & \dot{g}_{i,0,\ell-2} & \cdots & \dot{g}_{i,\nu,0} & 0 & \cdots & 0 \\
\vdots & \ddots & \ddots & \ddots & & \ddots & \ddots & \\
0 & \cdots & 0 & 0 & \dot{g}_{i,0,\ell-1} & \dot{g}_{i,0,\ell-2} & \cdots & \dot{g}_{i,\nu,0}
\end{bmatrix}
\end{equation}}
Notably, $\mathbf{\widetilde{G}}_i $ are $(N_f \cdot \ell) \times ((N_f+\nu+1)\cdot \ell-1)$ Toeplitz matrices, with a quite different structure compared to $\mathbf{\widetilde{H}}$. The reason is that, differently from the signal that is generated at the symbol rate, the noises are continuous time sources sampled at $\ell$ times the symbol rate after the convolution with the pulse responses of the filters. Therefore, for the noise sources, it is not possible to group the $\ell$ phases per symbol period, as is done to represent the convolution between $\tilde{h}(t)$ and the transmitted symbols $x_k$. Regarding the transceiver noise, $\widetilde{G}_\text{TRX}$ has the same structure shown in \eqref{eq:colored_noise_matrix}, but considering the pulse response $\tilde{g}_\text{TRX}(t)$ defined in \eqref{eq:trx_pulse_resp} to build it.

Finally, from \eqref{eq:colored_out_autocorr} and \eqref{eq:colored_inout_corr} we write the expression for the equalizer coefficients and the post-equalization noise variance:
\begin{align}
    \mathbf{\widetilde{w}} &= \tilde{R}_{x\mathbf{Y}}\cdot \tilde{R}_{\mathbf{YY}}^{-1} \\
            \sigma^2_\text{MMSEE,FLE} &= \bar{\mathcal{E}}_{\mathbf{x}} - \mathbf{\widetilde{w}} \cdot \tilde{R}_{\mathbf{Y}\mathbf{Y}} \cdot \mathbf{\widetilde{w}}^\dag
\end{align}
Then the unbiased SNR for FLE without noise whitening is obtained as follows.
\begin{equation}
    \label{eq:SNR_FLE_colored}
    {\mathrm{SNR}}_\text{MMSEE,FLE}=\frac{\bar{\mathcal{E}}_x}{{\sigma}^2_\text{MMSEE,FLE}}-1
\end{equation}


\subsection{SIMULATIONS}
\label{sec:simulations}

\begin{figure*}[htbp]
    \centering
    \includegraphics[width=\linewidth]{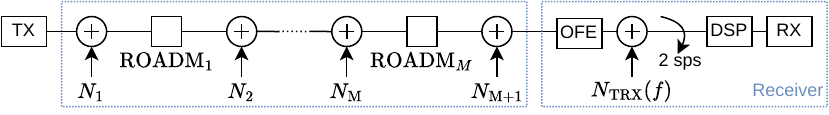}
    \caption{Reference setup for time-domain simulation with $M=10$ filtering stages.}
    \label{fig:timedomainsetup}
\end{figure*}

We performed the model validation through time-domain simulations, exploiting the software validated in \cite{PiloriSimul}. The adopted simulator implements least mean squares (LMS) algorithm to estimate the received symbol sequence, allowing to set the desired number of taps in the equalizer. We proceeded by comparing the SNR after equalization provided by the simulator with those computed with all the different models described in \autoref{sec:models}.

For validation purposes, we selected an arbitrary network setup containing ten identical filtering elements, replicating the filtering effect experienced by the signal when crossing ROADMs in network nodes. As anticipated, we adopted the optical filter model proposed in \cite{Pulikkaseril:11}, according to the filter shape of \eqref{eq:Pulikkaseril} setting the parameter $BW_\text{OTF}=11\;\text{GHz}$ and tuning the parameter $B_{ch}$ to obtain the desired equivalent 3-dB bandwidth of the filter cascade. We underline that the simulation allowed us to test the developed models on a scale unachievable in a laboratory environment, both in terms of number of elements and extreme filtering configurations.
To make meaningful comparisons of noise location effects noise sources can be placed before, after, or interspersed throughout the filter chain. \autoref{fig:timedomainsetup} visually details the scenario, in which the different noise sources can be enabled or disabled as needed, and the transmitted signal is root raised cosine (RRC) shaped, with symbol rate equal to 64 GBd, DP-QPSK modulation, and roll-off equal to 0.15.
Noise sources are tuned so that their combined effect produces an $\text{SNR}_\text{ASE}$ of $12\;\text{dB}$, while the receiver is always characterized by a $20\;\text{dB}$ $\text{SNR}_\text{TRX}$, with a white PSD. This choice was made to avoid the filtering penalty being predominated by transceiver noise.
In fact, if transceiver noise is the main impairment, a different displacement of ASE noise along the optical link does not bring to a significantly different performance.

For mathematical models and simulation results, performance is assessed by evaluating the $Q^2_\textit{dB}$ factor, as defined in \eqref{eq:Q_2}. The BER used in $Q$-factor evaluation is derived from the standard BER formula for DP-QPSK modulation:
\begin{equation}
    \text{BER} = \frac{1}{2}\text{erfc}\left( \sqrt{\frac{\text{SNR}_\text{EQ}}{2}} \right)
\end{equation}
where $\text{SNR}_\text{EQ}$ is the post-equalization SNR, and is computed from \eqref{eq:SNR_ZFE}, \eqref{eq:SNR_MMSEE}, \eqref{eq:SNR_FSE}, \eqref{eq:SNR_FLE}, and \eqref{eq:SNR_FLE_colored} for theoretical models; otherwise, it is provided directly by the time-domain simulator.

\begin{figure*}[htbp]
    \centering
    \includegraphics[width=\textwidth]{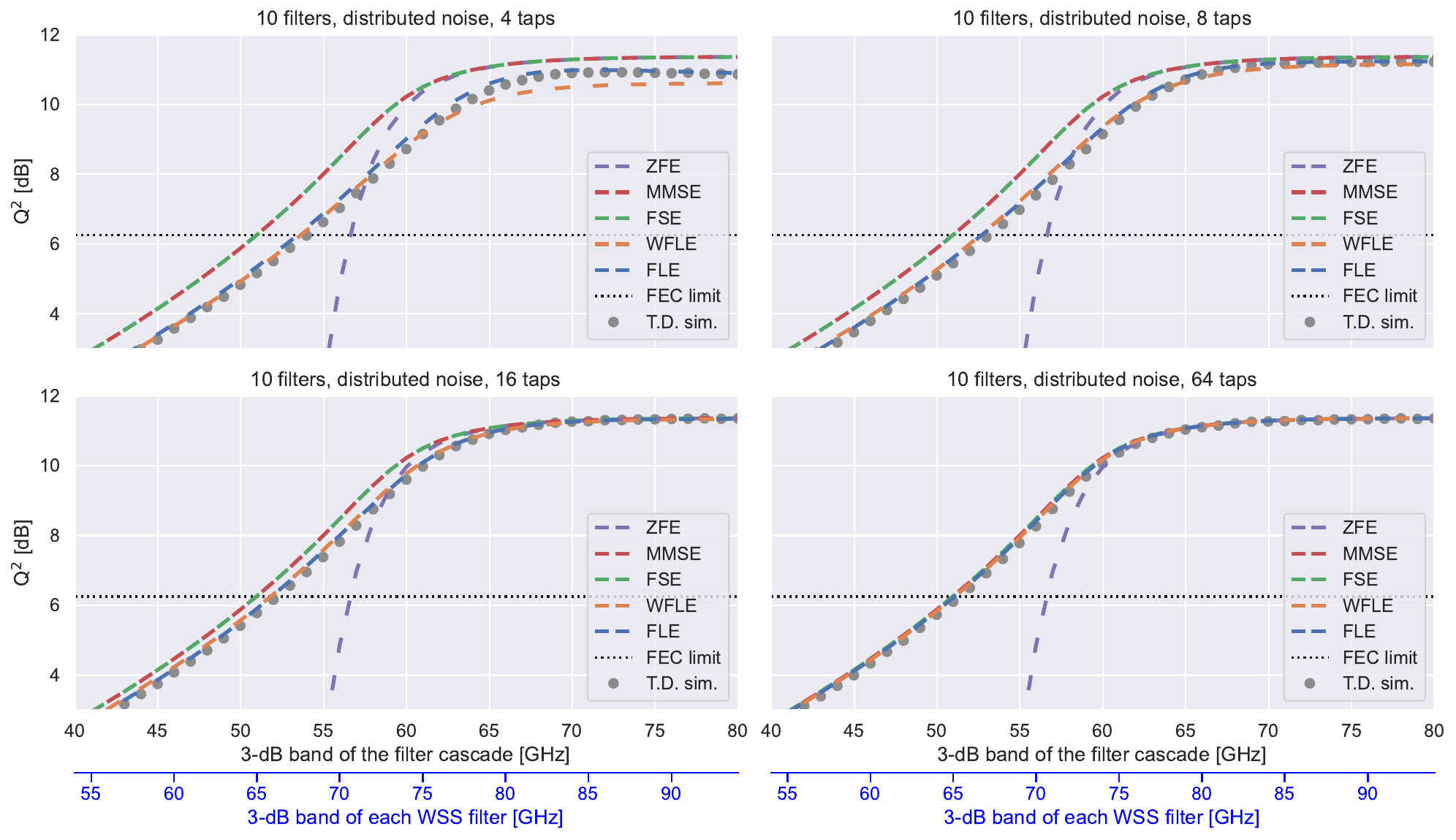}
    \caption{Equalizer model comparison and validation with time-domain simulations.}
    \label{fig:ssfm_taps_comparison}
\end{figure*}

The first validation step was a comparison of the different equalizer models with the result provided by the time-domain simulator, varying the number of taps used in the LMS algorithm and the total equivalent \mbox{3-dB} bandwidth of the filtering cascade. ASE noise has been uniformly distributed among noise sources. The maximum absolute errors for all the equalization models and simulation configurations are summarized in \autoref{tab:errors}. 

\autoref{fig:ssfm_taps_comparison} illustrates that an increase in the number of taps of the LMS algorithm allows an improvement in the estimation of the received signals, which in turn increases the $Q^2_\textit{dB}$ factor.
In particular, with $4$ or $8$ taps there is a non-negligible gap between finite and infinite equalizer models, since MMSEE and FSE cannot predict the limited-tap penalty. In contrast, increasing the LMS taps to $64$ makes all the results converge.
Additionally, MMSEE and FSE always provide the same result, as expected, since they are both based on mean square error. Indeed, in practical applications using commercial transceiver, FSE employing algorithms such as  the least mean squares or constant modulus algorithm converges to MMSEE \cite{CMA_vs_MMSEE}.

Regarding the WFLE and the FLE, we observe that, with 4 taps, there is a small difference between the two models. In particular, the WFLE is conservative with respect to time-domain simulation, with a maximum error of 0.41 dB, while the FLE shows to extremely accurately follow the time-domain simulation results, with a maximum error of 0.29 dB. Nevertheless, when the number of taps increases, as depicted for 8, 16 and 64 taps in \autoref{fig:ssfm_taps_comparison}, the WFLE almost overlaps to the FLE, with a difference in the estimation error lower than 0.05 dB. One explanation to consider is that the higher the number of taps, the closer the scenario to the case of infinite length equalizers, in which the presence of the whitening filter is no longer an approximation. Notably, the WFLE performs very reliable prediction also with 8 and 16 taps, with a maximum absolute error of 0.25 dB and 0.20 dB, where there is a non-negligible gap between the predictions of finite and infinite length Equalizer models. Therefore, in addition to the number of taps, it is also necessary to consider the total amount of noise and the way in which the channel pulse response is truncated. All these parameters have an impact on the dimension and on the values of the channel matrices introduced in Section \ref{sec:models}-\ref{sec:FLE}.

These results justify that the approach with WFLE in some cases was successfully adopted in past works, even when modeling finite length equalization \cite{miotto2025arxive, sdm_finite_taps}, but in general the FLE should be preferred, being more realistic and giving the best prediction performance, as summarized in \autoref{tab:errors}, where we provide also the RMSE, computed considering all the points for all the model-taps combinations. 

Lastly, the ZFE gives the worst performance prediction, when the equivalent bandwidth of the filter cascade is significantly lower than the signal symbol rate, due to strong noise enhancement when the filters are tight. Nevertheless, when the number of taps of the LMS algorithm is equal to 64, even the very light and simple ZFE gives a conservative and good assessment of the filtering penalty down to an equivalent filter cascade bandwidth of approximately the 90\% of the symbol rate, with a maximum absolute error of 0.87 dB. This observation is important because in a real-case scenario the overall filtering effect is not expected to be extreme, and commercial transceivers have a number of taps on the order of a few tens. Therefore, even the ZFE could be a good choice as a mathematical model, allowing fast penalty estimation and disaggregation, as anticipated in Section \ref{sec:models}-\ref{sec:infinite_equalizers}-\ref{sec:ZFE}. This result is pivotal for the development of an ONDT.

In addition, the computational complexity and the suitability for ONDT implementation must be carefully considered. Despite being much faster than time-domain simulation, the matrix-based FLE is the most computational-demanding among the considered equalization models.
Based on the FLE model described in Section \ref{sec:models}-\ref{sec:FLE}-\ref{sec:FLE_colored}, the overall computational cost of the FLE is approximately
{\footnotesize $O\left((N_f\ell)^2(N_f\!+\! \nu)+(M\!+\!2)(N_f\ell)^2((N_f\!+\! \nu \!+\!1)\ell \!-\!1)+ (N_f\ell)^3\right)$}, being $O((N_f\ell)^2(N_f+\nu))$ the cost of the matrix multiplication $\widetilde{H}\widetilde{H}^\dag$; $O((M+2)(N_f\ell)^2((N_f+\nu+1)\ell-1))$ the cost of the $M+2$ matrix multiplications $\widetilde{G}_i\widetilde{G}_i^\dag$ and $\widetilde{G}_\text{TRX}\widetilde{G}_\text{TRX}^\dag$; and $O((N_f\ell)^3)$ the computational cost to invert the $\tilde{R}_{\mathbf{YY}}$ matrix, which is the most expensive operation in order to compute $\mathrm{SNR}_\text{MMSEE,FLE}$. Overall, the computational complexity of the FLE is $\approx O(M(N_f\ell)^3)$. On the other hand, the WFLE does not require matrix multiplication for all the noise sources, therefore its computational complexity is $O((N_f\ell)^2(N_f+\nu)+(N_f\ell)^3) \approx O((N_f\ell)^3)$. Regarding the ZFE, MMSEE and FSE, they require to perform a single integration, which does not depend on the number of noise sources and is almost instantaneous with standard hardware. Therefore, it could be convenient to consider case-by-case which mathematical model fits better the application scenario.

\begin{figure}[htbp]
    \centering
    \includegraphics[width=1\linewidth]{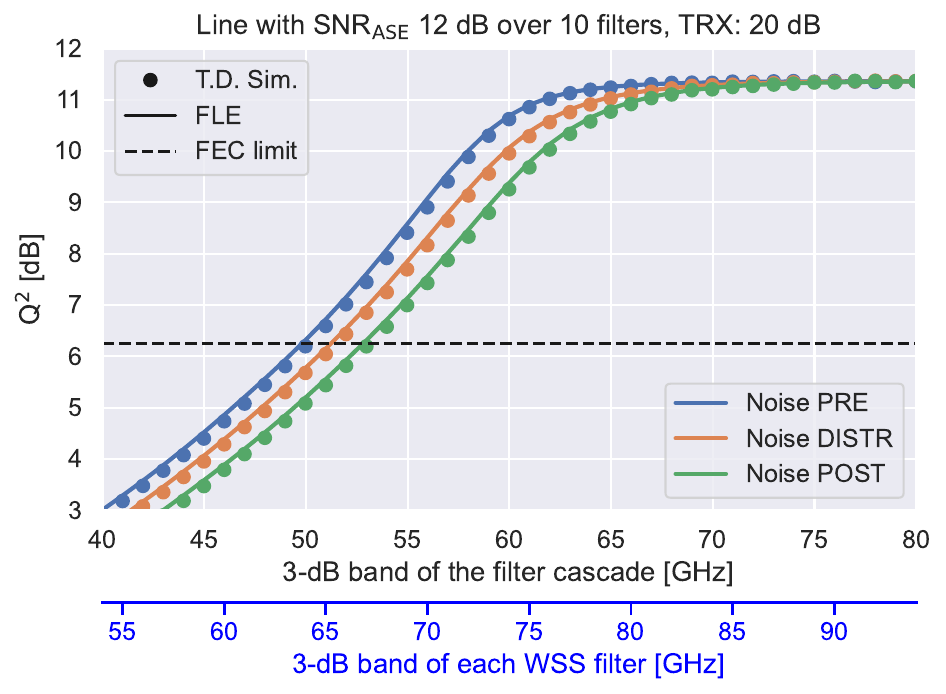}
    \caption{Comparison of the $Q^2_\textit{dB}$ evolution with different filter cascade bandwidths and noise positions. FLE equalizer with 32 taps.}    \label{fig:ssfm_pre_distr_post_comparison}
\end{figure}

\begin{table*}[ht]
\centering
\begin{tabular}{|c|cccc|cccc|cccc|}
\hline
 & \multicolumn{4}{c|}{Max. Abs. Error on $Q_{dB}^2$ [dB]} 
 & \multicolumn{4}{c|}{\makecell{Max. Abs. Error on $Q_{dB}^2$ [dB] when\\$B_{3dB} > 90\%$ of $R_s$}} 
 & \multicolumn{4}{c|}{RMSE [dB]}\\
 \hline
\backslashbox{Models}{Taps} & 4 & 8 & 16 & 64 & 4 & 8 & 16 & 64 & 4 & 8 & 16 & 64 \\
\hline
ZFE  & - & - & - & -  & - & - & - & 0.87 & - & - & - & - \\ 
MMSEE  & 1.56 & 1.15 & 0.69 & 0.21  & 1.56 & 1.15 & 0.69 & 0.18 & 0.92 & 0.65 & 0.37 & 0.12 \\ 
FSE  & 1.56 & 1.15 & 0.69 & 0.21  & 1.56 & 1.15 & 0.69 & 0.18 & 0.92 & 0.65 & 0.37 & 0.12 \\ 
WFLE  & 0.41 & 0.25 & 0.20 & 0.15  & 0.41 & 0.24 & 0.19 & 0.12 & 0.22 & 0.16 & 0.12 & 0.09 \\ 
FLE  & 0.29 & 0.20 & 0.17 & 0.15  & 0.29 & 0.19 & 0.15 & 0.12 & 0.16 & 0.11 & 0.10 & 0.09 \\  
\hline
\end{tabular}
\caption{Summary of the filtering penalty estimation absolute errors and RMSE for validation by means of time-domain simulation for different number of taps of the LMS equalizer. In the second group of columns, we considered only the cases in which the bandwidth of the cascade of the optical filters was larger than 90\% of the symbol rate.}
\label{tab:errors}
\end{table*}

In the second validation test, the number of LMS taps has been set to 32, and the ASE noise was injected differently by configuration: entirely at the first source in PRE, entirely at the last source in POST, and uniformly across sources in DISTR. In this way we explored how the interplay of filtering effect and noise distribution affects the system performance. In all three cases $\text{SNR}_\text{ASE}$ was kept constant and equal to 12 dB, as in the previous validation test, and the same for $\text{SNR}_\text{TRX}$, which was again set to 20 dB. 

The results of simulations in the time-domain have been compared with the mathematical model of the FLE, as reported in \autoref{fig:ssfm_pre_distr_post_comparison}. In this second test, it is also possible to observe a very precise agreement between the mathematical model and the points simulated using the LMS algorithm, with a maximum absolute error of 0.17 dB for PRE, 0.15 dB for DISTR, and 0.14 dB for POST.
Performance also depends on the relative position of noise and filtering elements. When noise is injected after the filter cascade, the SNR penalty is higher because equalization restores the signal while more strongly enhancing receiver-side noise. Conversely, noise injected closer to the transmitter passes through the full filter cascade, reducing equalizer-induced enhancement. Since receiver-side noise experiences the strongest filtering penalty, NLI can be conservatively modeled as AWGN injected at the receiver.
As shown in \autoref{fig:ssfm_pre_distr_post_comparison}, for the considered optical link, the minimum filter-cascade 3-dB bandwidth differs by about $4;\text{GHz}$ at the forward error correction (FEC) BER limit of $2\times10^{-2}$.

\section{Experimental Validation}

\begin{figure}[b]
    \centering
    \includegraphics[width=1\linewidth]{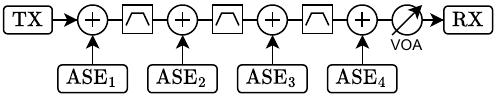}
    \caption{Experimental setup.}
    \label{fig:setup}
\end{figure}

\begin{figure*}
    \begin{center}
        \begin{subfigure}{0.24\textwidth}
            \includegraphics[width=1\linewidth]{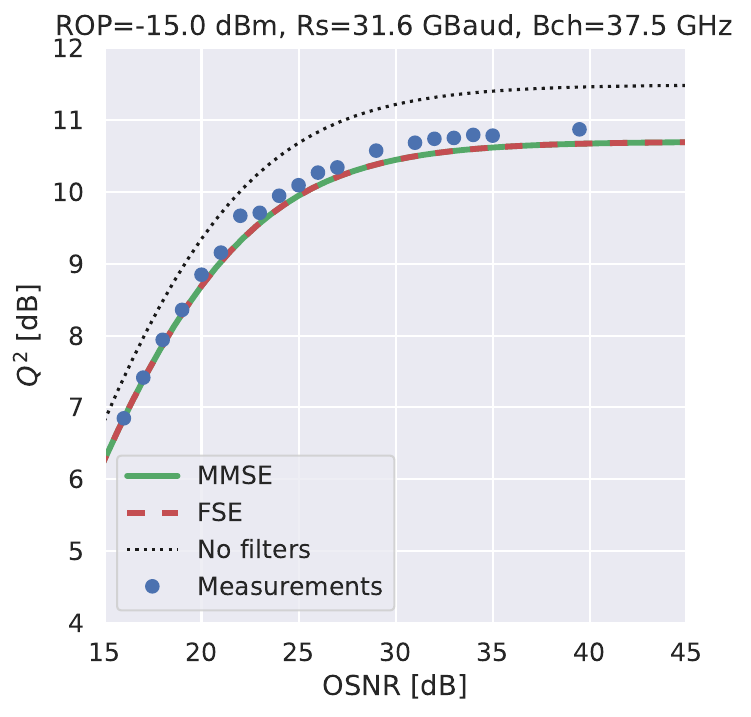}
            \caption{}
            \label{fig:1}
        \end{subfigure}
        \begin{subfigure}{0.24\textwidth}
            \includegraphics[width=1\linewidth]{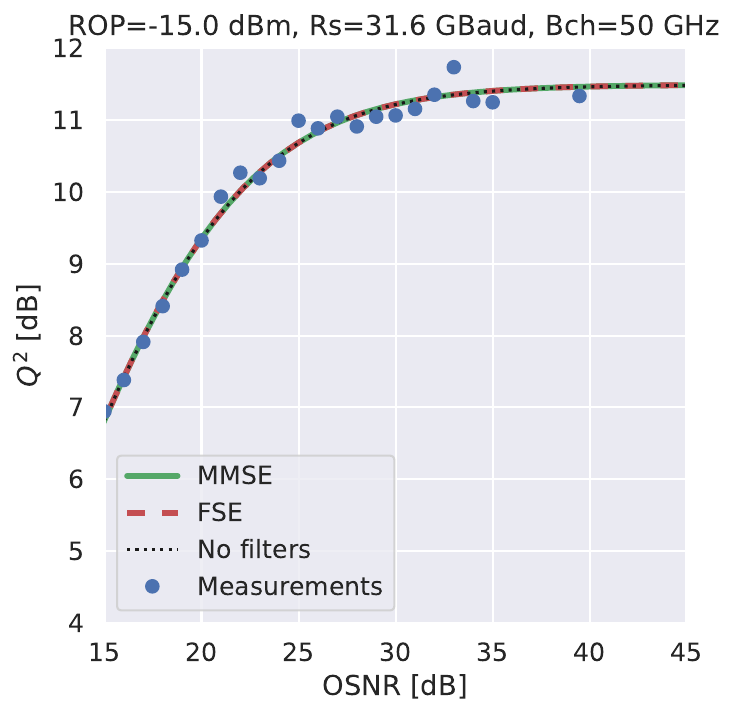}
            \caption{}
            \label{fig:2}
        \end{subfigure}
        \begin{subfigure}{0.24\textwidth}
            \includegraphics[width=1\linewidth]{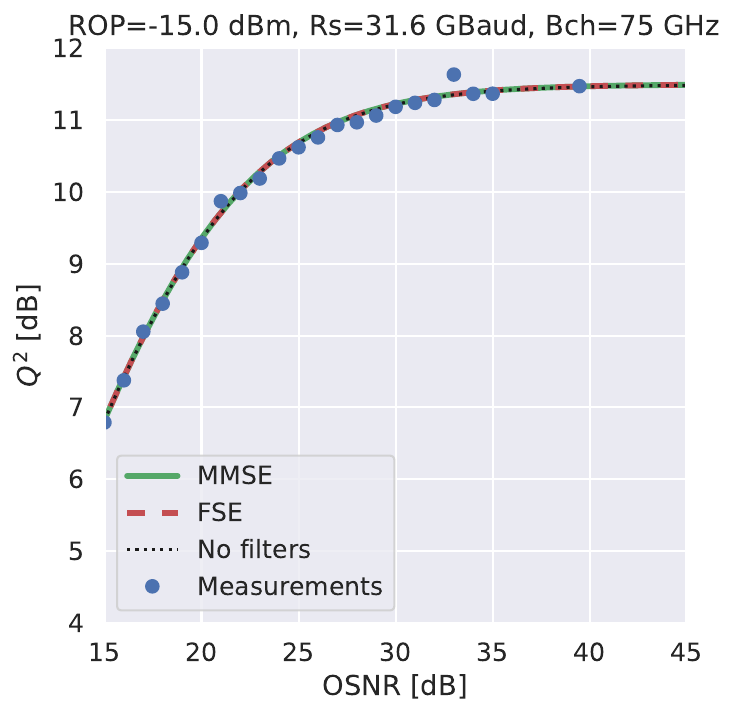}
            \caption{}
            \label{fig:3}
        \end{subfigure}
        \begin{subfigure}{0.24\textwidth}
            \includegraphics[width=1\linewidth]{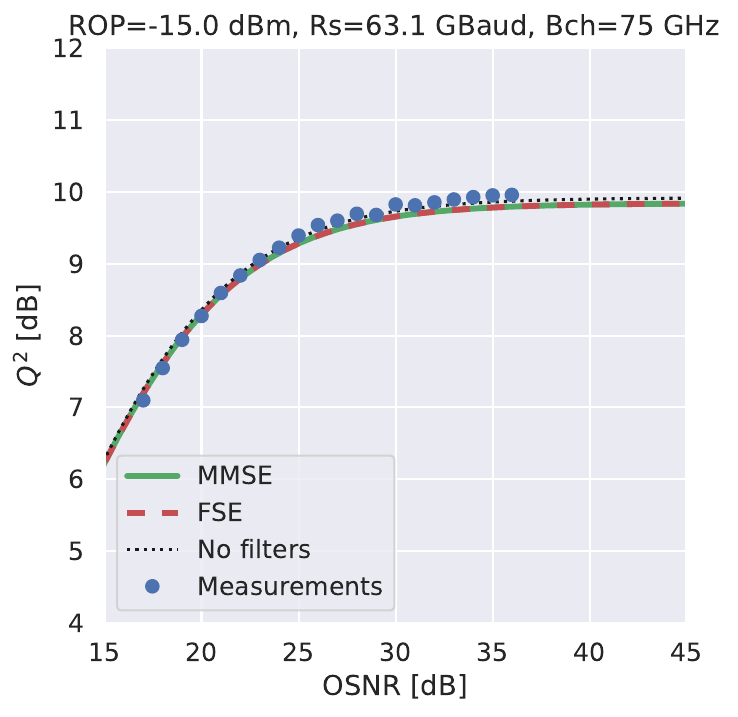}
            \caption{}
            \label{fig:4}
        \end{subfigure}

        \begin{subfigure}{0.24\textwidth}
            \includegraphics[width=1\linewidth]{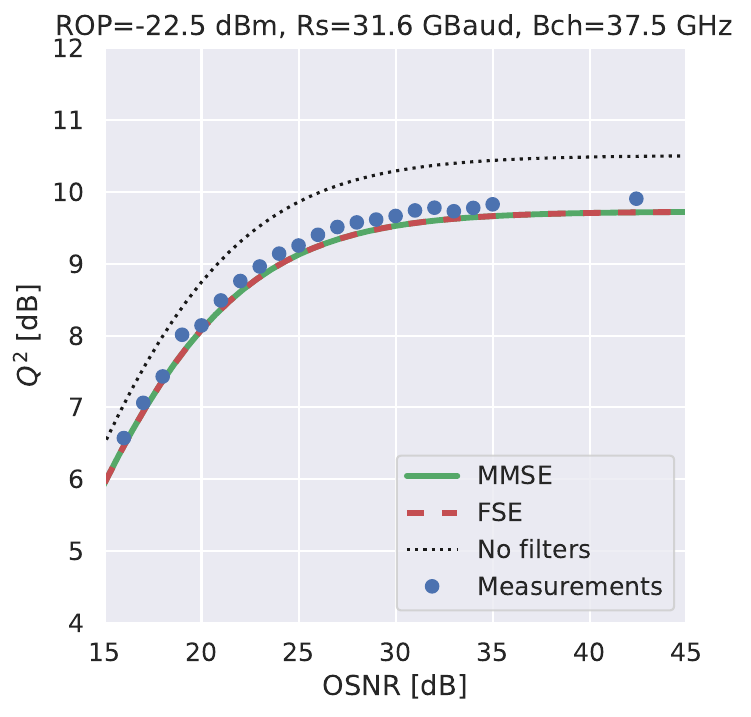}
            \caption{}
            \label{fig:5}
        \end{subfigure}
        \begin{subfigure}{0.24\textwidth}
            \includegraphics[width=1\linewidth]{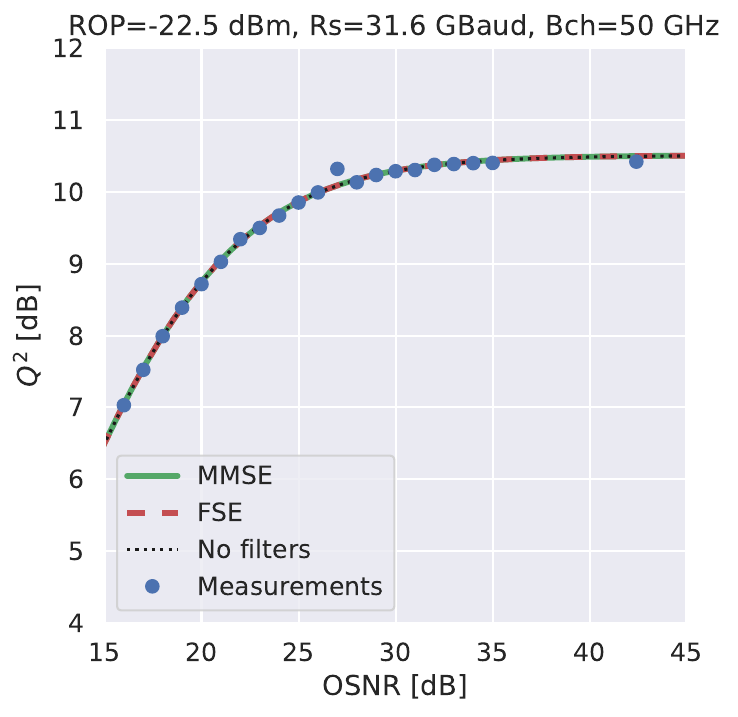}
            \caption{}
            \label{fig:6}
        \end{subfigure}
        \begin{subfigure}{0.24\textwidth}
            \includegraphics[width=1\linewidth]{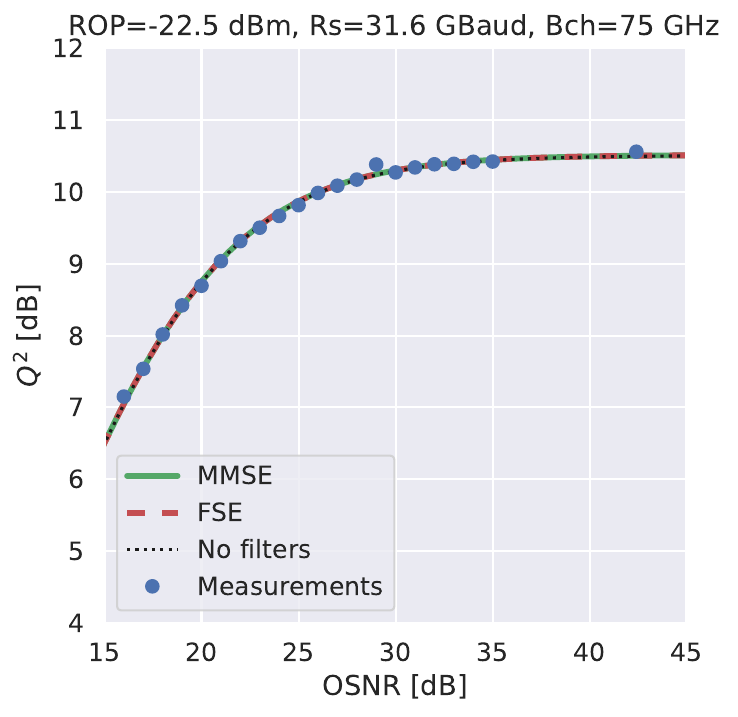}
            \caption{}
            \label{fig:7}
        \end{subfigure}
        \begin{subfigure}{0.24\textwidth}
            \includegraphics[width=1\linewidth]{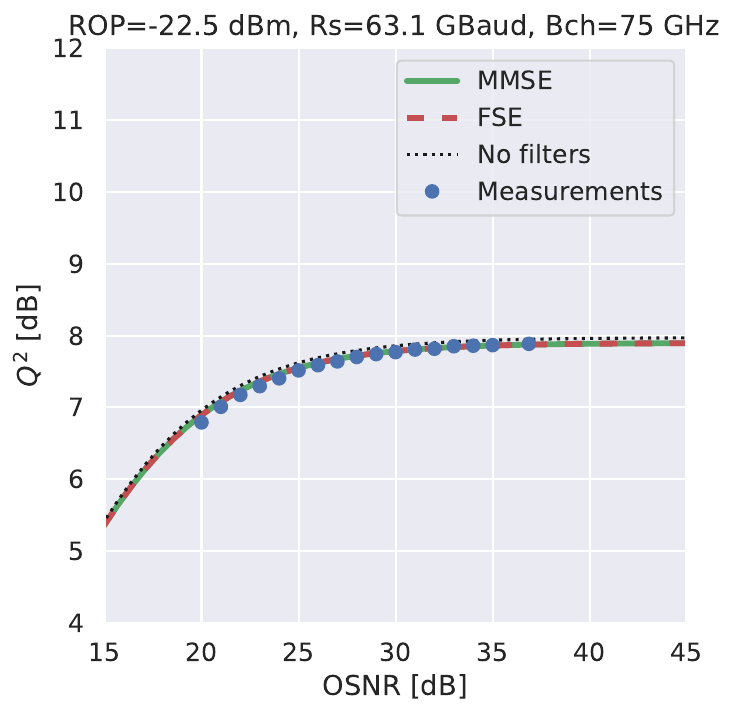}
            \caption{}
            \label{fig:8}
        \end{subfigure}

        \begin{subfigure}{0.24\textwidth}
            \includegraphics[width=1\linewidth]{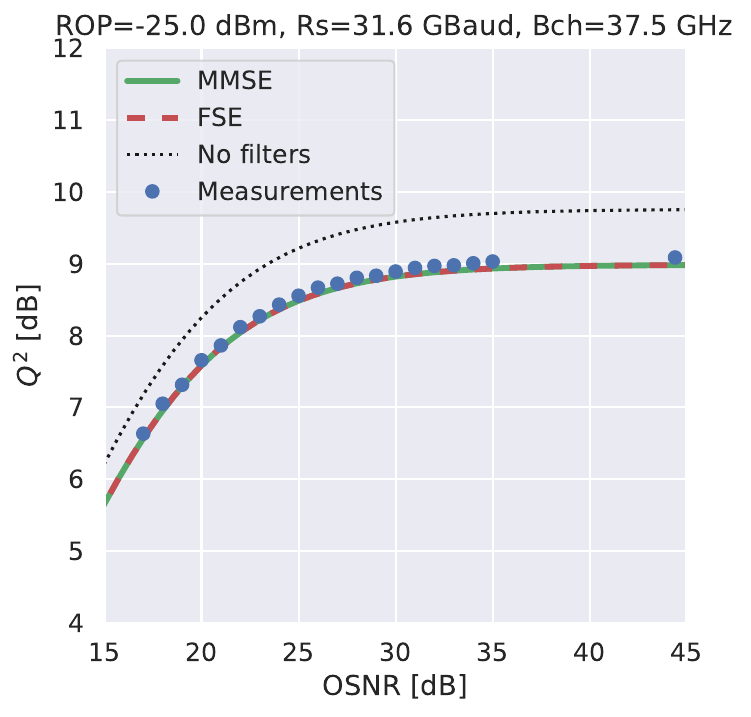}
            \caption{}
            \label{fig:9}
        \end{subfigure}
        \begin{subfigure}{0.24\textwidth}
            \includegraphics[width=1\linewidth]{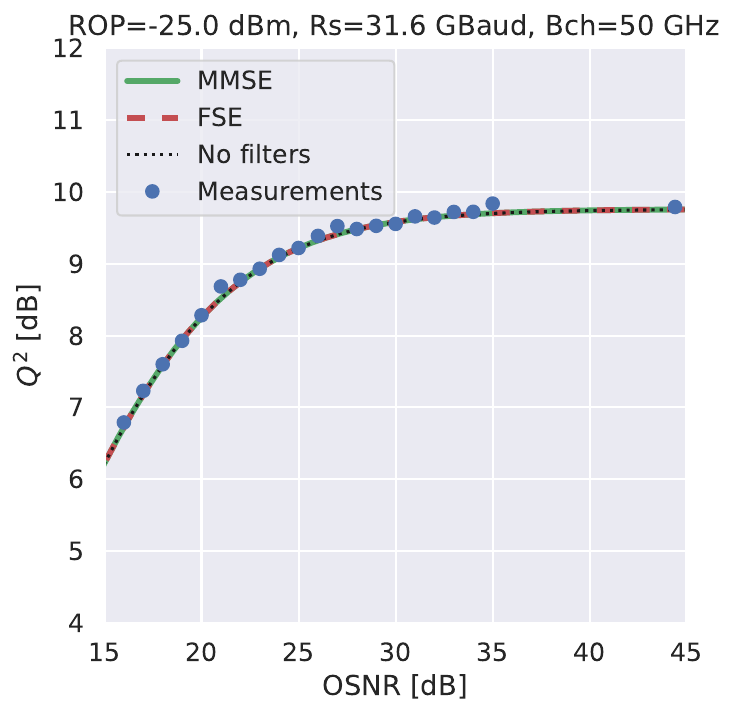}
            \caption{}
            \label{fig:10}
        \end{subfigure}
        \begin{subfigure}{0.24\textwidth}
            \includegraphics[width=1\linewidth]{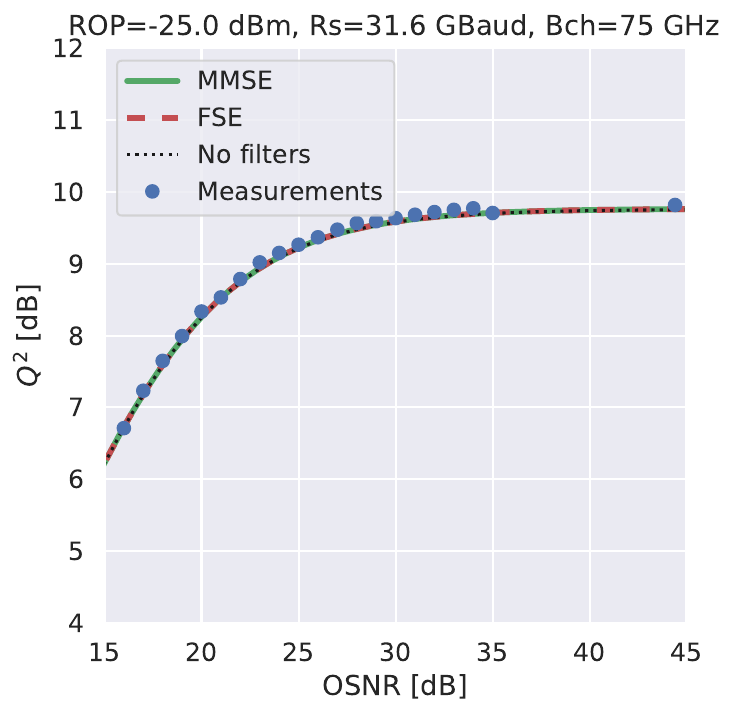}
            \caption{}
            \label{fig:11}
        \end{subfigure}
        \begin{subfigure}{0.24\textwidth}
            \includegraphics[width=1\linewidth]{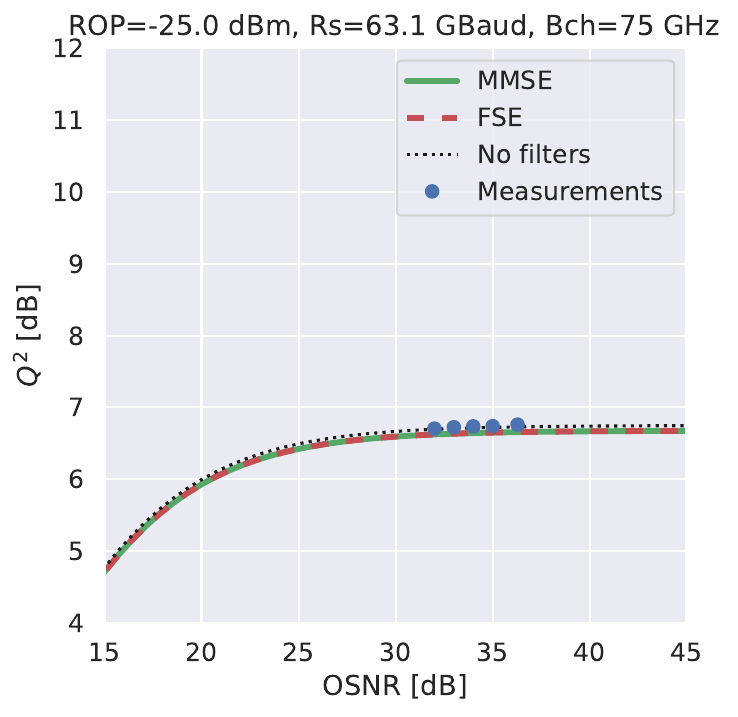}
            \caption{}
            \label{fig:12}
        \end{subfigure}
    \end{center}
    \caption{Experimental validation of MMSE and FSE equalizer models.}
    \label{fig:mosaic}
\end{figure*}

\subsection{Optical filters model}
\label{sec:filter_model}

The proposed models are general and can embed any optical filter shape. For experimental validation purpose, the spectral response of the optical filters was modeled following the approach in \cite{Pulikkaseril:11}, using the same parameters of \cite{ONDM2025}. Such optical filter model is well-suited to represent accurately the shape of filtering effect when commercial ROADMs are displaced in the optical link, and in general is preferable compared to supergaussian filter model. The filter transfer function amplitude is given by: 
\begin{equation}
    \label{eq:Pulikkaseril}
    S(f)=\frac{1}{2}\sigma\sqrt{2\pi}
\left [ 
\mathrm{erf}\left ( \frac{\frac{B_\text{ch}}{2}-f}{\sigma\sqrt2} \right )
-
\mathrm{erf}\left ( \frac{-\frac{B_\text{ch}}{2}-f}{\sigma\sqrt2} \right )
\right ] 
\end{equation}
Where $B_\text{ch}$ is the channel bandwidth and $\sigma = \frac{BW_\text{OTF}}{2\sqrt{2ln2}}$.
The OSA-measured traces of the filter spectra were fitted by adjusting the $BW_{\text{OTF}}$ parameter. A conservative fitting approach was adopted, resulting in modeled spectra that are slightly narrower than the experimental data near the edges.

\subsection{Comparison between models and experiment}

The experimental validation of the models developed in Section \ref{sec:models} is provided in Figure \ref{fig:mosaic}. The measurements are the same used in \cite{ONDM2025} to validate the ZFE model. In this work we perform the validation of MMSE and FSE, since the ZFE model of Section \ref{sec:ZFE} is equivalent to the one of \cite{ONDM2025}. Regarding the FLE, since the actual number of taps of the Equalizer embedded in the commercial transceiver was not known, we leave the experimental validation of the model of Section \ref{sec:FLE} to be included in further experimental campaigns.

The experimental setup used in \cite{ONDM2025} to investigate filtering effects is recalled in Figure \ref{fig:setup} ant it was composed by a cascade of three ROADM filters aligned to the channel under test. Filter bandwidths were reconfigurable from 37.5 to 75 GHz in 12.5 GHz steps, creating a symmetric dual-sided filtering condition. A booster amplifier compensated for insertion losses, maintaining a constant launch power of 0 dBm. Experiments were conducted using DP-16QAM at 200 and 400 Gbps with 31.6 and 63.1 Gbaud signals. ASE noise was independently controlled and calibrated across four sources to ensure $\mathrm{SNR}_{\mathrm{ASE}}$ is equal between the four noise sources when measured at the receiver. BER was measured at different received optical power (ROP) levels and $\mathrm{SNR}_{\mathrm{ASE}}$ values, while an optical spectrum analyzer (OSA) was used to capture the signal spectrum and measure OSNR.

Figure \ref{fig:mosaic} is organized as follows: from left to right the optical filters bandwidth changes and also the symbol rate of the transmitted signal. From top to bottom the received optical power changes from -15 dBm to -25 dBm. The plots are presented in terms of the well-known Q factor \cite{Freude:2012}, which is derived from the measured BER values using the following equations:
\begin{align}
    \label{eq:Q}
    Q &= \sqrt{2} \cdot \operatorname{erfcinv}(2 \cdot \mathrm{BER})\\
    \label{eq:Q_2}
    Q^2_{\mathrm{dB}} &= 10 \cdot \log_{10}(Q^2)
\end{align}
For the analytical model, the electrical SNR is calculated for MMSE and FSE equalizers using respectively Equations \ref{eq:SNR_MMSEE} and \ref{eq:SNR_FSE}. The bit error rate (BER) is then determined using the standard theoretical expression \cite{Carena:2012}:
\begin{equation}
    \label{eq:ber}
    \mathrm{BER} = \frac{3}{8} \cdot \operatorname{erfc} \left( \sqrt{\frac{1}{10} \cdot \mathrm{SNR}} \right)
\end{equation}
This analysis considers DP-16QAM modulation, where SNR is evaluated for the various filter configurations under investigation. The resulting BER values are then converted to $Q^2_{\mathrm{dB}}$ using Equations \ref{eq:Q} and \ref{eq:Q_2}.

Additionally, theoretical reference curves (shown as dashed lines in Figure \ref{fig:mosaic} represent the ideal case without any filter-induced penalty. These curves are derived exploiting the transceiver characterization, as done in \cite{ONDM2025} and using Equation \ref{eq:ber} to evaluate the corresponding BER.

Overall, the analytical models align well with the experimental data, maintaining a slightly conservative margin, particularly under stronger filtering conditions. The models' validity is first confirmed by the overlap between the analytical curves and experimental points when filter bandwidths are significantly wider than the symbol rate: consistent with the theoretical no-filtering reference. The conservative margin visible in Figures \ref{fig:1}, \ref{fig:5} and \ref{fig:9} arises from the conservative filter modeling approach described in Section \ref{sec:filter_model}. Additionally, discrepancies between the experimental results and the model can be attributed to the use of ideal, infinite-length equalizers in the model, which differ from the practical implementation in the actual transceiver. Furthermore, Figures \ref{fig:4}, \ref{fig:8} and \ref{fig:12} display a lower Q factor, which is attributed to the higher bitrate configuration. This setup introduces a more significant transceiver penalty, as already stated in \cite{ONDM2025}.

Another interesting consideration comes out by comparing the MMSE and FSE equalizers performance, which are expressed by identical superposed curves in Figure \ref{fig:mosaic}. As stated in the derivation in Section \ref{sec:models}, FSE is the more realistic version of MMSE equalizer, since it does not rely on matched filtering in continuous domain. Since also FSE is based on mean square error as optimization metric, it is expected to have similar performance. The advantage of FSE in commercial transceivers is that it allows an improvement in sensitivity to sampling-phase errors, which is important to limit noise enhancement that would occur if using MMSE and ZFE equalizers. Since in the derivation we did not consider sampling-phase deviation, the overlapping of MMSE and FSE performance curves is correct and expected.

\section{Conclusions}
This work presented a discrete-time analytical framework for evaluating the combined effect of cascaded optical filtering, distributed ASE noise, transceiver noise, and receiver equalization in coherent optical links. The model starts from a general optical-link abstraction and derives post-equalization SNR expressions for the MFB, ZFE, MMSE, FSE, and finite-length equalizers. A central aspect of the formulation is the explicit treatment of colored noise: ASE contributions injected at different points of the link are filtered differently, while transceiver noise is added at the receiver and can be represented through a measurement-based SNR model.

The time-domain simulations confirm the validity of the derivations over severe filtering conditions and show the role of equalizer length. Infinite-length MMSE and FSE models correctly describe the asymptotic behavior, whereas the finite-length model captures the penalty observed when the LMS equalizer has a limited number of taps. The comparison also shows that the whitened finite-length approximation can be accurate in several practical cases, while the full colored-noise FLE remains the most faithful model when implementation details must be represented more precisely. ZFE provides a simpler and conservative estimate when the filtering cascade is not excessively tight, which is useful for low-complexity quality-of-transmission evaluation.

The experimental validation with commercial transceivers and ROADM filters further supports the proposed approach. The MMSE and FSE curves reproduce the measured Q-factor trends with good agreement and a conservative margin mainly associated with the filter-fitting strategy and with practical DSP implementation limits. The overlap between MMSE and FSE predictions in the absence of sampling-phase errors is consistent with the theoretical derivation, while the higher penalty observed in more demanding bitrate configurations is captured through the transceiver characterization.

Overall, the results show that filtering penalties cannot be represented only by a scalar bandwidth loss: they depend on filter shape, noise-source position, receiver noise, and equalizer architecture. The proposed framework provides a tractable way to include all these effects in system-level planning tools and optical-network digital twins. Future work should extend the experimental validation of finite-length equalizers by characterizing the actual DSP tap configuration of commercial transceivers and by testing a wider set of filter shapes, sampling conditions, and lightpath configurations.

\balance
\section*{Acknowledgements}
This work has been partially funded by the EU - Next Generation EU under the Italian NRRP, Mission 4, Component 2, Investment 1.3, CUP E13C22001870001, partnership on “Telecommunications of the Future” (PE00000001 - program “RESTART”) and by the Telecom Infraproject.

\printbibliography

\end{document}